\documentclass[sigconf]{acmart}
\AtBeginDocument{%
  \providecommand\BibTeX{{%
    \normalfont B\kern-0.5em{\scshape i\kern-0.25em b}\kern-0.8em\TeX}}}

\setcopyright{acmcopyright}
\copyrightyear{2018}
\acmYear{2018}
\acmDOI{XXXXXXX.XXXXXXX}

\acmConference[Conference acronym 'XX]{Make sure to enter the correct
  conference title from your rights confirmation emai}{June 03--05,
  2018}{Woodstock, NY}
\acmPrice{15.00}
\acmISBN{978-1-4503-XXXX-X/18/06}

\usepackage{subcaption}
\usepackage{multirow}
\usepackage{color}
\usepackage{pgfplots}
\pgfplotsset{width=10cm,compat=1.9}
\usetikzlibrary{pgfplots.groupplots}
\usepackage{booktabs}
\usepackage{arydshln}
\usepackage{multicol}
\usepackage{rotating}
\usepackage{collcell}
\usepackage{makecell}

\usepackage{pgfkeys}
    \newenvironment{customlegend}[1][]{%
        \begingroup
        \csname pgfplots@init@cleared@structures\endcsname
        \pgfplotsset{#1}%
    }{%
        \csname pgfplots@createlegend\endcsname
        \endgroup
    }%
    \def\addlegendimage{\csname pgfplots@addlegendimage\endcsname}

\acmSubmissionID{609}

\begin{document}

\title[ONCE]{ONCE: Boosting Content-based Recommendation with Both Open- and Closed-source Large Language Models}
% \subtitle{ User Interest Generation,
% User Profile Tracking, and News Feature Enhancement}
\author{Qijiong Liu}
\email{liu@qijiong.work}
\affiliation{%
  \institution{The Hong Kong Polytechnic University}
  \city{Hong Kong}
  \country{China}
}

\author{Nuo Chen}
\email{pleviumtan@toki.waseda.jp}
\affiliation{%
  \institution{Waseda University}
  \city{Tokyo}
  \country{Japan}}

\author{Tetsuya Sakai}
\email{tetsuyasakai@acm.org}
\affiliation{%
  \institution{Waseda University}
  \city{Tokyo}
  \country{Japan}}

\author{Xiao-Ming Wu}
\authornote{Corresponding author.}
\email{xiao-ming.wu@polyu.edu.hk}
\affiliation{%
  \institution{The Hong Kong Polytechnic University}
  \city{Hong Kong}
  \country{China}
}

\renewcommand{\shortauthors}{Liu, et al.}


\begin{abstract}
    Personalized content-based recommender systems have become indispensable tools for users to navigate through the vast amount of content available on platforms like daily news websites and book recommendation services. However, existing recommenders face significant challenges in understanding the content of items.
    Large language models (LLMs), which possess deep semantic comprehension and extensive knowledge from pretraining, have proven to be effective in various natural language processing tasks.
    In this study, we explore the potential of leveraging both open- and closed-source LLMs to enhance content-based recommendation. 
    With open-source LLMs, we utilize their deep layers as content encoders, enriching the representation of content at the embedding level. For closed-source LLMs, we employ prompting techniques to enrich the training data at the token level.
    Through comprehensive experiments, we demonstrate the high effectiveness of both types of LLMs and show the synergistic relationship between them. Notably, we observed a significant relative improvement of up to 19.32\% compared to existing state-of-the-art recommendation models. These findings highlight the immense potential of both open- and closed-source of LLMs in enhancing content-based recommendation systems. We will make our code and LLM-generated data available\footnote{\url{https://github.com/Jyonn/ONCE}} for other researchers to reproduce our results.

\end{abstract}





\maketitle
% \definecolor{bookblue}{rgb}{0.48, 0.52, 0.71}
% \definecolor{bookgreen}{rgb}{0.59, 0.73, 0.58}
% \definecolor{bookorange}{rgb}{0.90, 0.63, 0.40}
\definecolor{bookblue}{rgb}{0.38, 0.42, 0.57}
\definecolor{bookgreen}{rgb}{0.47, 0.58, 0.46}
\definecolor{bookorange}{rgb}{0.72, 0.50, 0.32}
\newcommand{\lionking}{``\textit{\textcolor{bookblue}{The Lion King}}''}
\newcommand{\alrassan}{``\textit{\textcolor{bookorange}{The Lions of Al-Rassan}}''}
\newcommand{\summer}{``\textit{\textcolor{bookgreen}{The Summer Tree}}''}

% \begin{figure}[t]
%     \centering
%     \setlength\tabcolsep{10pt}
%     \begin{tabular}{cc}
%     \begin{subfigure}{0.30\linewidth}
%         \resizebox{\linewidth}{!}{
%             \includegraphics{images-v2/Paradigm-Open.pdf}
%         }
%         \caption{}
%     \end{subfigure} &
%     \begin{subfigure}{0.52\linewidth}
%         \resizebox{\linewidth}{!}{
%             \includegraphics{images-v2/Paradigm-Closed.pdf}
%         }
%         \caption{}
%     \end{subfigure}
%     \end{tabular}

%     \caption{\label{fig:paradigm} Illustration of the integration of (a) open-source and (b) closed-source large language models with recommendation model.}
% \end{figure}

\begin{figure}[t]
    \centering
    \includegraphics[width=.9\linewidth]{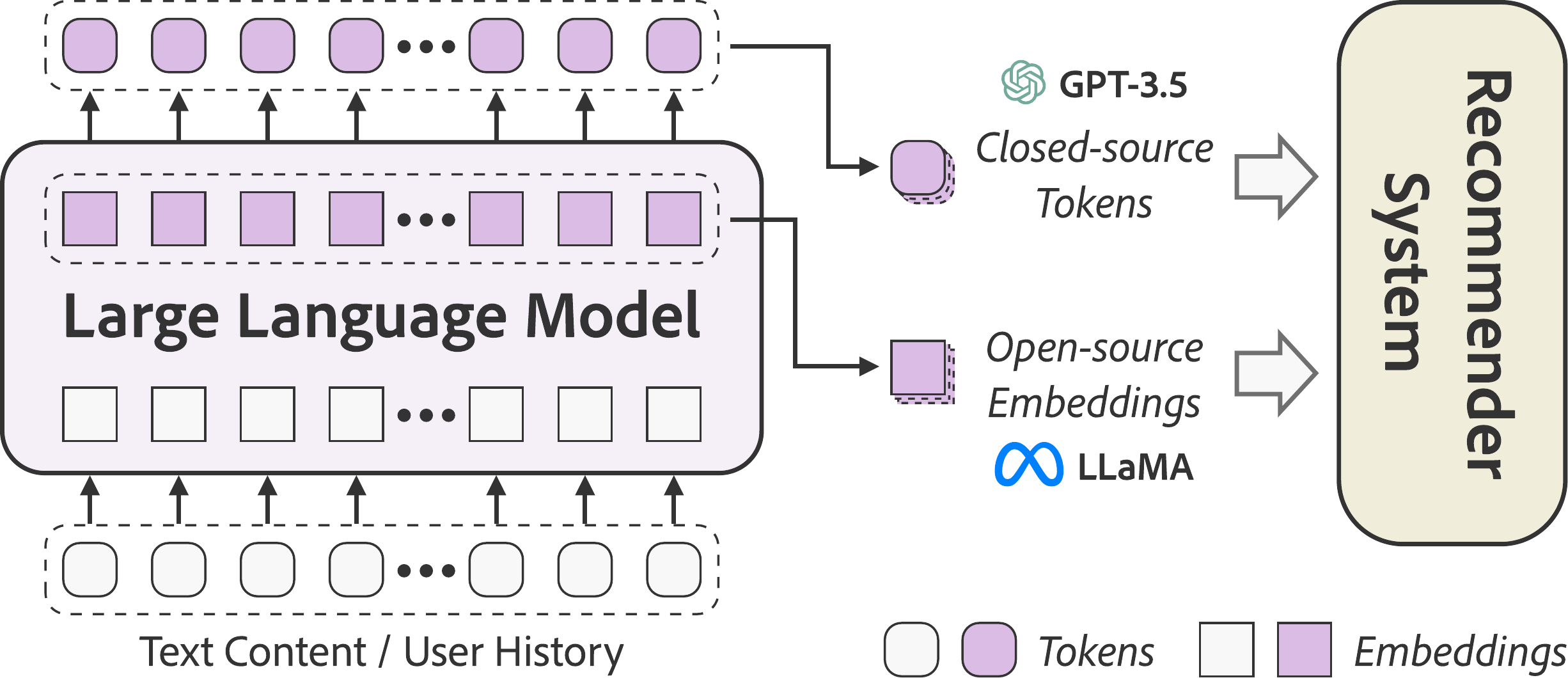}
    \caption{\label{fig:paradigm} 
    %Two types of use of large language models for content-based recommendation.
    Employing two different types of LLMs for content-based recommendations.
    }
\end{figure}

\begin{figure*}[t]
    \centering
    \includegraphics[width=.7\linewidth]{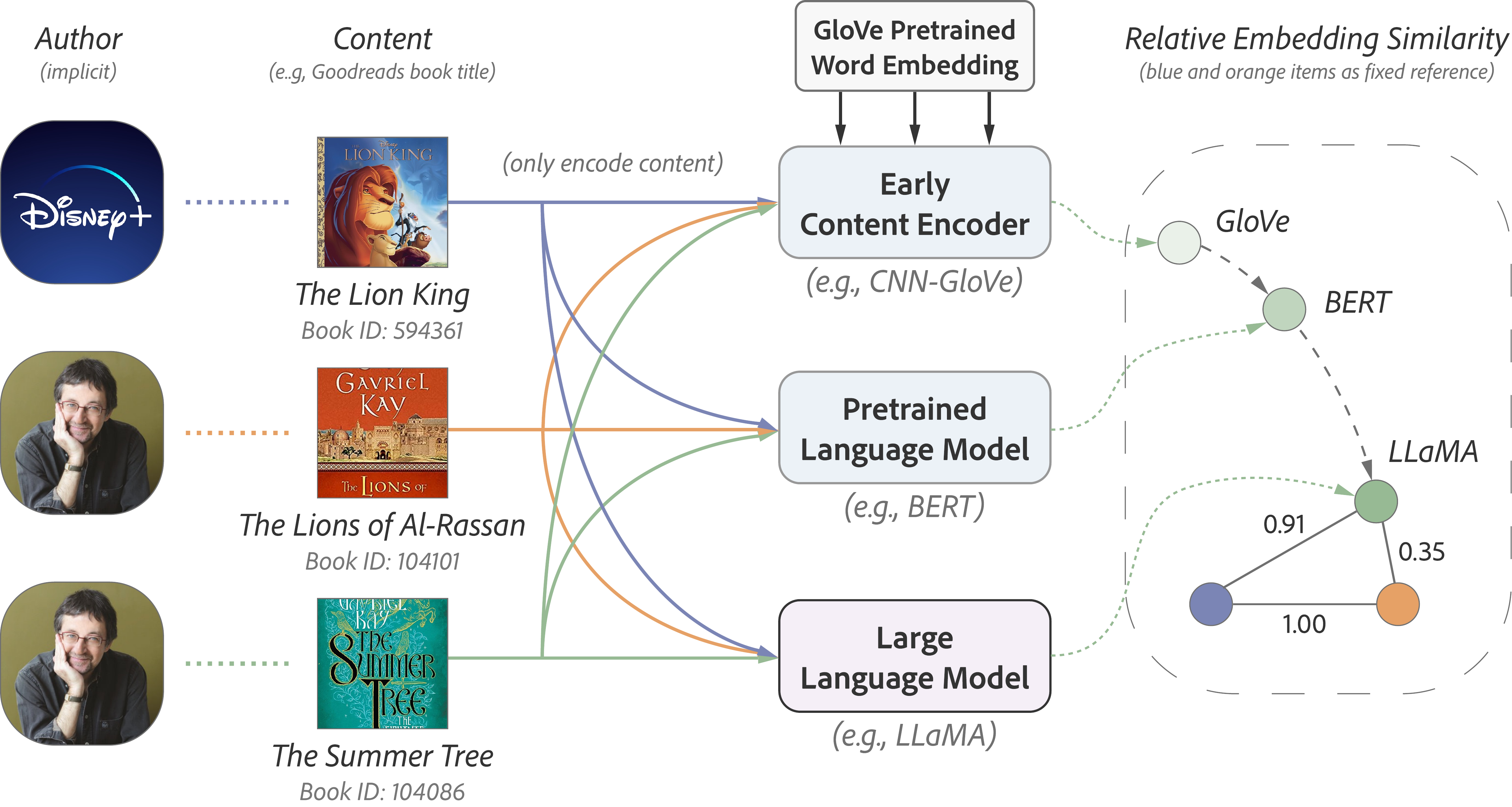}
    \caption{\label{fig:introduction} Comparison of content encoders used for content-based recommendation. 
     To illustrate the similarity among the three books,  we employ relative distance to align the three different embedding spaces. First, we compute the cosine similarities $s_{i, j}$ between each pair of books. Then, we calculate their relative distances using $d_{i, j} = (1-s_{i, j})/(1-s_{blue, orange})$, i.e., by fixing the similarity between \lionking{} and \alrassan{} as 1. This approach allows for a direct comparison of their similarities across the three distinct embedding spaces. It's important to note that a shorter distance indicates a greater similarity.}
\end{figure*}

\section{Introduction}

% Content-based recommendation systems are vital for information retrieval and user experience, delivering personalized and relevant content to users based on their preferences.
% Examples of such systems include Google News~\footnote{~\url{https://news.google.com/}} and Goodreads~\footnote{~\url{https://www.goodreads.com/}}, which provide personalized news and book recommendations.
% As the volume of digital content increases, improving content-based recommendation techniques becomes essential to meet users' demands for accurate and relevant suggestions.

Content-based recommender systems analyze the content and properties of items (e.g., articles, movies, books, or products) to deliver relevant and personalized recommendations to users. Some instances of such systems are Google News\footnote{~\url{https://news.google.com/}}, which offers recommendations for news articles, and Goodreads\footnote{~\url{https://www.goodreads.com/}}, which provides recommendations for books. With the rapid expansion of digital content, it becomes increasingly essential to improve content-based recommendation techniques in order to meet users' expectations for precise and pertinent recommendations.

%Content-based recommendation systems leverage textual information by constructing content encoders to capture semantic features. In the past, shallow networks such as convolutional neural networks (CNNs) were commonly used~\cite{naml,nrms,lstur}, initialized with pre-trained word representations such as GloVe~\cite{glove}. In recent years, recommendation models have harnessed the power of pretrained language models (PLMs) to extract more comprehensive semantic information~\cite{plmnr}. Despite these advancements, these methods still face limitations in effectively incorporating external knowledge.

The core component of content-based recommender systems is the \emph{content encoder}, which is used for encoding the textual information of items in order to capture semantic features. In the past, recommendation models~\cite{naml,nrms,lstur} commonly utilized convolutional neural networks (CNNs) as content encoders, typically initialized with pre-trained word representations such as GloVe~\cite{pennington2014glove}. 
In recent years, recommendation methods~\cite{plmnr} have made use of pretrained language models (PLMs) based on the Transformer architecture~\cite{attention} to extract more comprehensive semantic information. Despite these advancements, existing methods still struggle to fully comprehend the content of items.

\begin{figure*}[ht]
    \centering
    \includegraphics[width=.9\linewidth]{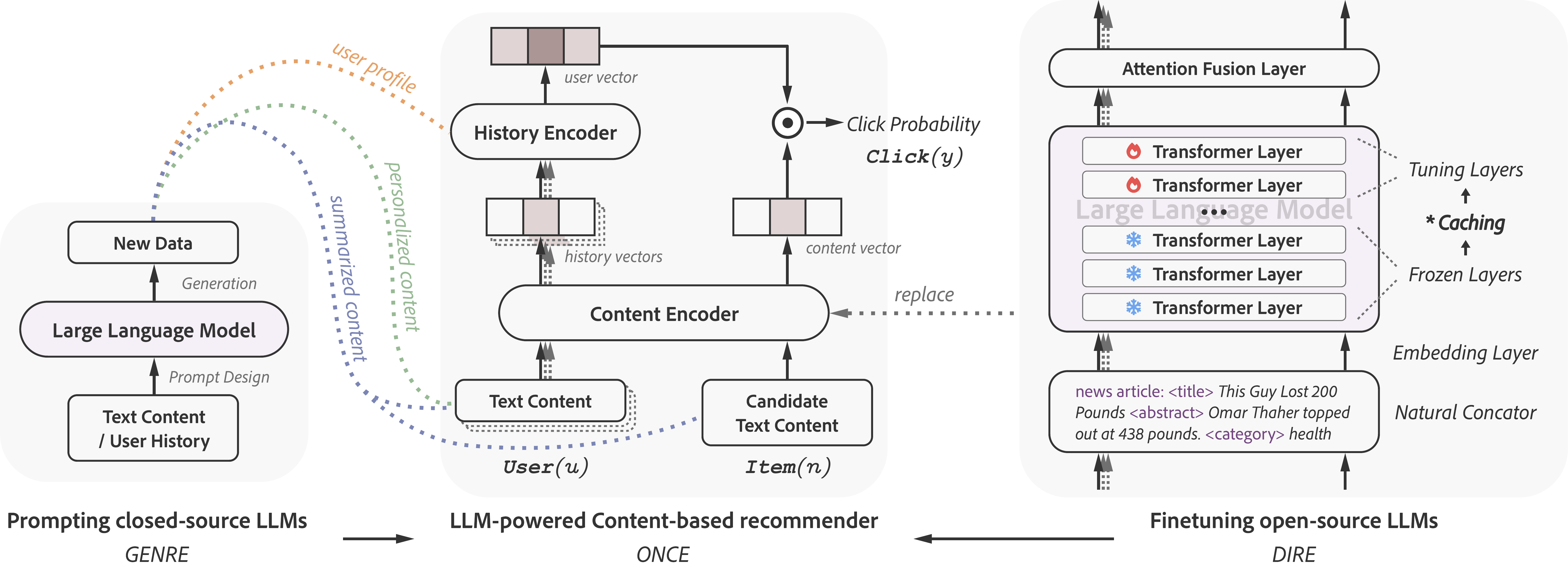}
    \caption{\label{fig:once} An overview of our proposed ONCE framework, designed to enhance content-based recommendations by finetuning open-source LLMs (DIRE) and employing prompts for closed-source LLMs (GENRE).}
\end{figure*}

%Here is an example demonstrating the significance of external knowledge.
To illustrate the limitations of previous content encoders, we present an example in ~\autoref{fig:introduction}.
We chose three books from the Goodreads dataset: \lionking{}, a novel adapted from a Disney animated movie, and the historical fantasy novels \alrassan{} and \summer{}, both authored by Guy Gavriel Kay, belonging to a distinct category. We use different encoders to encoder the titles of these books and visualize their relative similarities in the embedding space. The results reveal that early content encoders relying on pretrained word embeddings struggle at the \emph{word level}, failing to recognize crucial terms like ``Al-Rassan'' and resulting in erroneously high similarity between ``The Lion King'' and ``The Lions of Al-Rassan''.
Similarly, small-scale pretrained language models (PLMs) face challenges at the \emph{content level}. Constrained by their pretraining data lacking relevant knowledge and their limited representation dimensions (e.g., 768), they are unable to fully grasp the content of ``The Lions of Al-Rassan'' and ``The Summer Tree'' and accurately perceive their similarity, leading to outcomes similar to the early content encoders.

%When encoding these titles with various content encoders, ~\autoref{fig:introduction} reveals relative similarities of their representations.
%Traditional approaches using pretrained word embeddings struggle at the \textbf{word level}, failing to recognize terms like ``Al-Rassan.''
%Consequently, the similarity between ``The Lion King'' and ``The Lions of Al-Rassan'' is erroneously high compared to their similarity with ``The Summer Tree.'' Similarly, small-scale pretrained language models encounter challenges at the \textbf{content level}, limited by the lack of relevant knowledge in their pretraining data and their constrained representation dimensions (e.g., 768). As a result, the similarity outcomes do not significantly differ from traditional methods.

% Most recently, we have witnessed the emergence of the large language model (LLM) era, with the likes of ChatGPT\footnote{~\url{https://chat.openai.com}} and LLaMA~\cite{llama} leading the way.
% These models boast billions of parameters and are trained on datasets containing trillions of tokens.
% With each token represented in thousands of dimensions, they can store a wealth of information.
% The ``emergent abilities''~\cite{wei2022emergent} of these large language models have opened up new horizons for recommendation systems, offering unprecedented possibilities.
% Their advanced language understanding capabilities empower them to deliver more contextually relevant and personalized recommendations, significantly enhancing user satisfaction and engagement in the process.

Such limitations can be overcome by the emerging large language models (LLMs), with the likes of closed-source ChatGPT\footnote{~\url{https://chat.openai.com}} and open-source LLaMA~\cite{llama} leading the way.
These models possess billions of parameters and are trained on datasets containing trillions of tokens.
With each token represented in thousands of dimensions, they can store an extensive amount of information. In contrast to small PLMs like BERT, these LLMs demonstrate remarkable ``emergent abilities''~\cite{wei2022emergent} in terms of advanced language comprehension and generation capabilities, making it possible to deliver more contextually relevant and personalized recommendations. 
When we use ChatGPT to inquire about a book, it showcases its enriched knowledge at the content level by providing detailed information such as the author, publication date, and subject matter.
In \autoref{fig:introduction}, we initially prompted LLaMA to generate concise descriptions of the three books solely based on the title information, and then employed it to encode these descriptions and obtain the corresponding representations\footnote{It is important to note that in our experiments in Section \ref{sec:exp}, we used LLaMA as a content encoder without any prompting.}. The results clearly indicate that the representations generated by LLaMA accurately reflect the similarity in content between the three books: the similarity between \alrassan{} and \summer{} is higher than their similarity with \lionking{}.

In this paper, we investigate the possibility of  enhancing content-based recommendation by leveraging both \textbf{O}pe\textbf{N}- and \textbf{C}los\textbf{E}d-source (\textbf{ONCE}) LLMs. As depicted in~\autoref{fig:paradigm}, our approach ONCE adopts different strategies for each type of LLMs. For open-source LLMs like LLaMA, we employ a \emph{\textbf{di}scriminative} \textbf{re}commendation approach named \textbf{DIRE}, reminiscent of the PLM-NR~\cite{plmnr} method, by replacing the original content encoder with the LLM. This enables us to extract content representations and fine-tune the model specifically for recommendation tasks, ultimately enhancing user modeling and content understanding. Conversely, for closed-source LLMs like GPT-3.5, where we only have access to token outputs, we propose a \emph{\textbf{gen}erative} \textbf{re}commendation approach named \textbf{GENRE}. 
By devising various prompting strategies, we enrich the available training data and acquire more informative textual and user features, which contribute to improved performance in downstream recommendation tasks.

We conducted extensive evaluations using two well-established content recommendation benchmarks: MIND~\cite{mind} and Goodreads~\cite{goodreads}.
Our main objective was to thoroughly assess the impact of both open-source and closed-source LLMs on content-based recommendation models, focusing on recommendation quality and training efficiency. The results of our study demonstrate that both open- and closed-source LLMs are highly effective, especially the former. Through the process of finetuning LLaMA, we consistently observed enhancements of more than 10 percentage points compared to existing state-of-the-art recommendation models. Additionally, we discovered a complementary relationship between open- and closed-source LLMs. Specifically, the enriched data generated by ChatGPT substantially accelerated the efficiency of finetuning LLaMA while simultaneously enhancing recommendation quality.
%This synergy between the two types of LLMs underscores their potential for elevating content-based recommendation systems. 
Our findings highlight the immense potential of both types of LLMs in enhancing content-based recommendation systems.

\section{Overview}

\subsection{Content-based Recommendation}

Before delving into the details of our proposed method, we first introduce basic notations and formally define the content-based recommendation task.
Let $\mathcal{N}$ represent the set of contents, where each content $n \in \mathcal{N}$ is characterized by a diverse feature set, such as title, category, or description, in various recommendation scenarios.
Similarly, let $\mathcal{U}$ denote the set of users, where each user $u \in \mathcal{U}$ maintains a history of browsed content denoted as $h^{(u)}$.
Additionally, $\mathcal{D}$ corresponds to the set of click data, with each click $d \in \mathcal{D}$ represented as a tuple $(u, n, y)$, indicating whether user $u$ clicked on content $n$ with label $y \in {0,1}$.
The objective of content-based recommendation is to infer the user's interest in a candidate content.

A content-based recommendation model typically consists of three core modules: a content encoder, a history encoder, and an interaction module.
The content encoder is responsible for encoding the multiple features of each content, consolidating them into a unified $d$-dimensional content vector $\mathbf{v}_n$.
On top of the content encoder, the history encoder generates a unified $d$-dimensional user vector $\mathbf{v}_u$ based on the sequence of browsed content vectors.

Finally, the interaction module aims to identify the positive sample that best aligns with the user vector $\mathbf{v}_u$ among multiple candidate content vectors $\mathbf{V}_c = [\mathbf{v}_c^{(1)}, ..., \mathbf{v}_c^{(k+1)}]$, where $k$ represents the number of negative samples.
This process can be viewed as a classification problem.

\subsection{Enhancing Content-based Recommendation with Open- and Closed-source LLMs (ONCE)}

Large language models, endowed with deep semantic understanding and comprehensive knowledge acquired from pretraining, have exhibited proficiency across a multitude of natural language processing tasks. In this paper, we introduce the ONCE framework, which leverages both open-source and closed-source LLMs to enhance content-based recommendations. 
% Taking advantage of the feasible and complementary aspects of tuning open-source LLMs and prompting closed-source LLMs, we present ONCE, an innovative amalgamation of both large model types.
As highlighted in Touvron et al.~\cite{touvron2023llama}, there remains a discernible gap between open-source models, encompassing approximately 10 billion parameters, and the closed-source GPT-3.5, an expansive entity boasting over 175 billion parameters.
Our ONCE framework capitalizes on the strengths of both types and constructs a more robust recommendation system.
We initiate the process by utilizing the closed-source LLM through prompting, enhancing the dataset from various perspectives, following our designed generative recomendation framework (GENRE).
This infusion of external knowledge, a facet not readily accessible to open-source models, ensues.
Subsequently, we propose a discriminative recommendation framework (DIRE) to harness the deep layers of the open-source LLM as content encoders, thereby amplifying content representations.
\newcommand{\sep}{$\langle$sep$\rangle$}
\newcommand{\cls}{$\langle$cls$\rangle$}
\newcommand{\feature}{$\langle$\textit{feature}$\rangle$}

\section{DIRE: Finetuning Open-source LLMs}

% \input{figures-v2/open-source}

%\subsection{Overview}

Integrating open-source language models as content encoders is a straightforward and widely adopted method in content-based recommendation~\cite{liu2022prec,plmnr}.
Notably, PLM-NR~\cite{plmnr} employs small-scale pretrained language models (PLMs, e.g., BERT~\cite{bert}) to replace original news encoders and finetunes on the recommendation task.

The success of this approach relies on two factors: 1) the knowledge inherent in the pretrained language models (including model size and pretraining data quality), and 2) the finetuning strategy.
As discussed earlier, we have already highlighted the advantages of large language models in content understanding and user modeling, addressing the first factor.
In this section, we propose discriminative recommendation framework, namely \textbf{DIRE}, and explore how to leverage open-source large language models to further enhance recommendation performance by considering the second factor.

\subsection{Network Architecture}

As depicted in~\autoref{fig:once}, we seamlessly incorporate the open-source large language model and an attention fusion layer into the content-based recommendation framework.

% \begin{figure*}[t]
%     \centering
%     \includegraphics[width=.8\linewidth]{images-v2/Closed-Source.pdf}
%     \caption{\label{fig:closed_source} Prompting closed-source GPT-3.5 as data augmenter.}
% \end{figure*}

\begin{figure*}[t]
  \centering
  \begin{subfigure}[b]{0.35\textwidth}
    \centering
    \includegraphics[width=\linewidth]{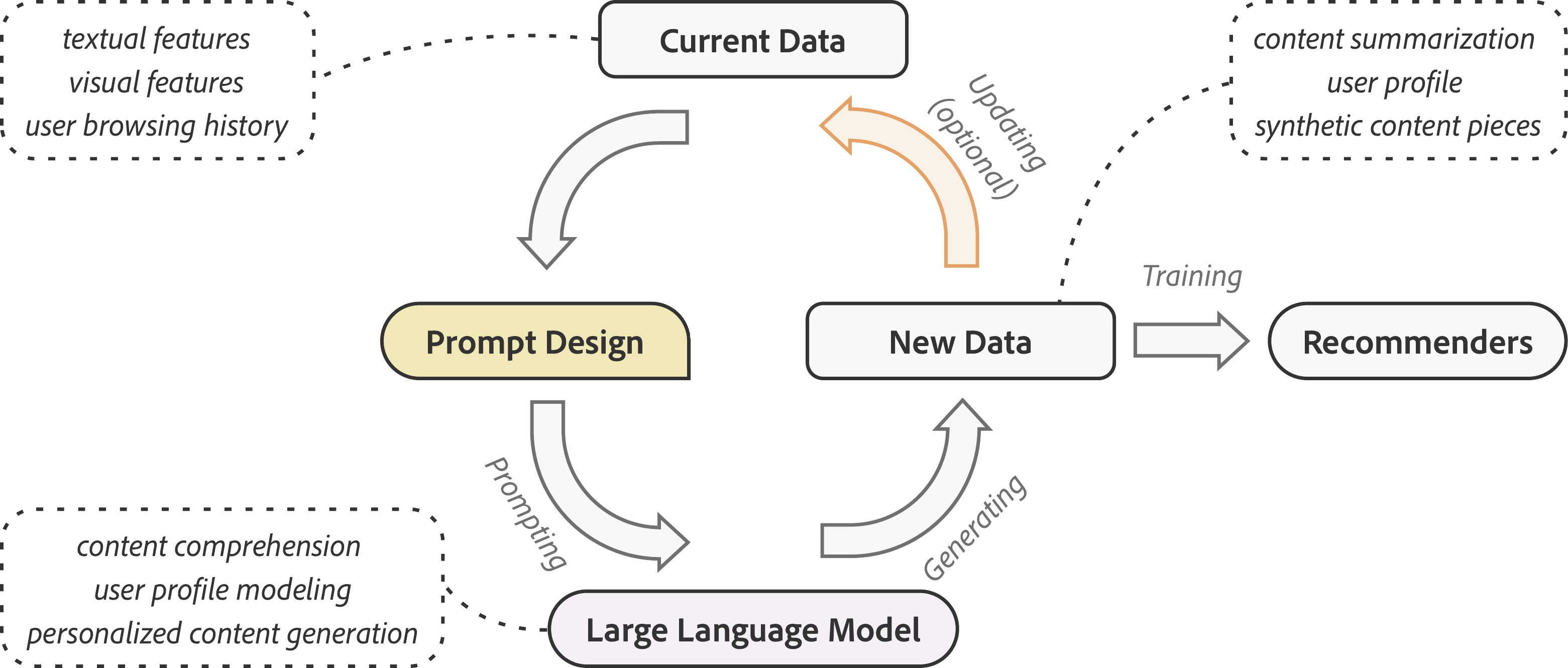}
    \caption{\label{fig:genre}}
  \end{subfigure}%
  \hspace{0.05\textwidth} 
  \begin{subfigure}[b]{0.55\textwidth}
    \centering
    \includegraphics[width=\linewidth]{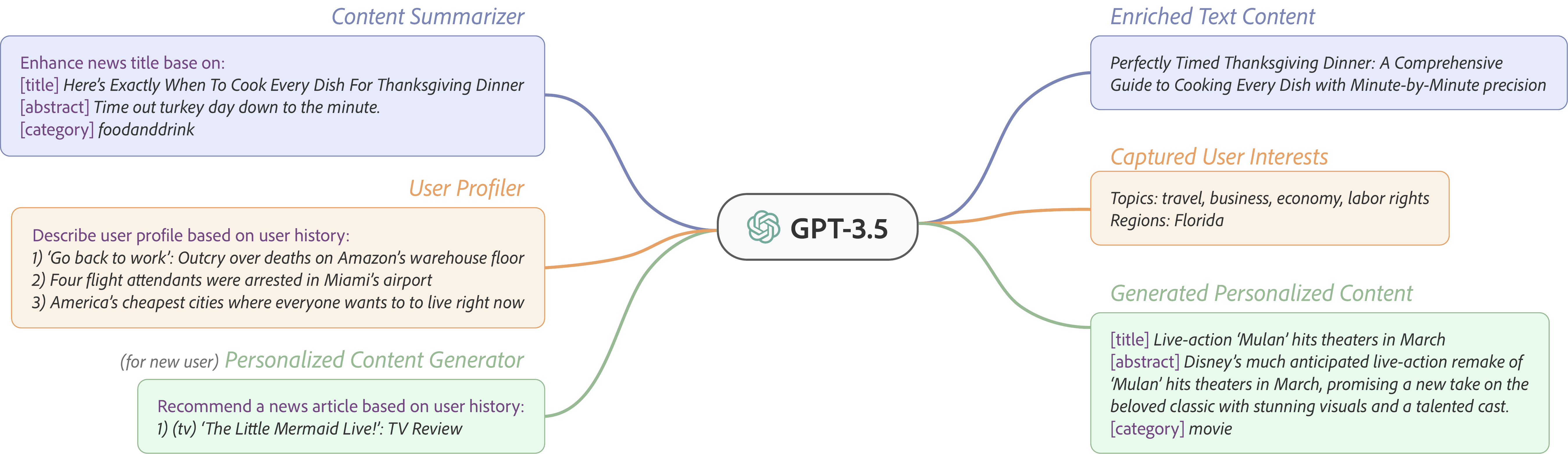}
    \caption{\label{fig:prompts}}
  \end{subfigure}
  \caption{\label{fig:closed_source} (a) Our proposed GENRE framework. (b) Prompting closed-source GPT-3.5 as data augmenter.}
\end{figure*}

\textbf{Embedding Layer.}
In contrast to the approach taken by smaller-scale PLMs like BERT, which utilize specific tokens (e.g., \cls{}, \sep{}) to segment distinct fields, we adopt natural language templates for concatenation.
For instance, consider a news content $n$ containing attributes such as title, abstract, and category features.
As illustrated in~\autoref{fig:once}, We introduce the label ``\textit{news article:}'' at the outset of the sequence, while each feature is prefixed with ``\feature{}''.
This procedure transforms the multi-field content into a cohesive individual sequence $\mathbf{s}$ of length $l$.
We refer to this technique as the ``\textbf{Natural Concator}''.
Following this, we make use of pretrained token embeddings provided by the LLM to map discrete text sequence into a continuous embedding space of dimensionality $d^n$, denoted as:
\begin{equation}
    \mathbf{E^0} = EmbeddingLayer\left(\mathbf{s}\right) \in \mathbb{R}^{l \times d^n}.
\end{equation}

\textbf{Transformer Decoder.}
The design of the LLM (or LLaMA) is based on the Transformer architecture~\cite{attention}, incorporating multiple tiers of Transformer Layers.
This configuration is intricately interconnected, with the output hidden state from each layer feeding into the input of the next layer, denoted as:
\begin{equation}
    \mathbf{E^i} = TransformerLayer\left(\mathbf{E^{i-1}}\right) \in \mathbb{R}^{l \times d^n}, i \in \{1, ..., H\},
\end{equation}
where $H$ represents the number of Transformer Layers.

\textbf{Attention Fusion Layer.}
To combine the sequential hidden states from the last layer into a single cohesive content representation, we employ the attention fusion layer, following a similar approach as used in PLM-NR~\cite{plmnr}.
Specifically, we begin by mapping the high-dimensional hidden states from a large space of dimensionality $d^n$ to a smaller $d$-dimensional space (where $d^n \gg d$), defined by:
\begin{equation}
    \mathbf{Z} = \mathbf{E^i} \mathbf{W} + \mathbf{b} \in \mathbb{R}^{l \times d},
\end{equation}
where $\mathbf{W} \in \mathbb{R}^{d^n \times d}$ and $\mathbf{b} \in \mathbb{R}^{d}$ are the learnable parameters of the linear transformation.
Next, we utilize the additive attention mechanism~\cite{bahdanau2014neural} to further condense the reduced representation into a unified representation $\mathbf{z}$, defined by:
\begin{equation}
    \mathbf{z} = Attention\left(\mathbf{Z}\right) \in \mathbb{R}^{d},
\end{equation}
which will be fed into the user modeling module or interaction module for further personalized recommendation.

\subsection{Finetuning Strategy}

\textbf{Partial Freezing and Caching.}
Running large language models incurs significant computational demands due to their expansive collection of transformer layers and associated parameters.
Given that the lower layers of the LLM tend to possess a more generalized and less task-specific nature, we opt to keep these layer parameters fixed.
Instead, we exclusively fine-tune the uppermost $k$ layers, where $H \gg k$.
Furthermore, we adopt a caching strategy wherein we precompute and store the hidden states from the lower layers for all contents within the dataset (potentially numbering in the thousands) prior to fine-tuning.
For instance, in the case of the LLaMA-7B model comprising 32 layers, only the top 2 layers are subjected to fine-tuning.
This caching process substantially mitigates computational costs, reducing the LLM's computation load to a mere $2/32 \approx 6\%$ of the original cost.

\textbf{Parameter-Efficient Tuning.} Low-Rank Adaptation (LoRA)~\cite{hu2021lora} introduces trainable rank decomposition matrices into pretrained model layers, notably slashing the necessary trainable parameters for downstream tasks.
This outperforms traditional fine-tuning, drastically reducing parameters, sometimes by a factor of 10,000.
Here, we apply LoRA to the unfrozen Transformer layers, which are the most parameter-intensive components of the model.
We also test finetuning without LoRA, employing \textit{distinct learning rates} for the pretrained Transformer layers and other model components.
Further elaboration is available in the Experiments section.

\section{GENRE: Prompting Closed-Source LLMs}\label{sec:genre}

%\subsection{Overview}

Large language models differ significantly from previous models like BERT~\cite{bert} in terms of their emergent abilities~\cite{wei2022emergent} such as strong text comprehension and language generation capabilities, having resulted in a paradigm shift from the traditional pretrain-finetune approach to the prompting-based approach.
Previous studies~\cite{lin2023can} have found that using closed-source LLMs directly as recommenders without finetuning (completely bypassing conventional recommendation systems), using methods like prompts~\cite{wang2023zero,liu2023chatgpt} or in-context learning~\cite{dai2023uncovering}, only matches the performance of basic matrix factorization~\cite{koren2009matrix} methods or even random recommendations.
This falls short when compared to modern attention-based approaches.

To overcome this, we propose a generative recommendation framework, namely \textbf{GENRE}, as shown in~\autoref{fig:genre}: leveraging closed-source LLMs (specifically, GPT-3.5) to augment data, aiming to enhance their performance on downstream conventional recommendation models.
More precisely, the workflow consists of the following four steps.
\textbf{1) Prompting}: create prompts or instructions to harness the capability of a LLM for data generation for diverse objectives.
\textbf{2) Generating}: the LLM generates new knowledge and data based on the designed prompts.
% \textbf{3) Training}: leverage the generated data to train news recommendation models.
\textbf{3) Updating} \textit{(optional)}: use the generated data to update the current data for the next round of prompting and generation.
\textbf{4) Training}: leverage the generated data to train news recommendation models.
If the updating step is performed, we name it as ``\textit{Chain-based Generation}'', otherwise, we name it as ``\textit{One-pass Generation}''.

% \section{Proposed Framework: GENRE}

% \subsection{Overview}

% \autoref{fig:idea} illustrate the our proposed GENRE framework for LLM-powered generative news recommendation, which consists of the following four steps. 1) Prompting: create prompts or instructions to harness the capability of a LLM for data generation for diverse objectives. 2) Generating: the LLM generates new knowledge and data based on the designed prompts. 3) Updating: use the LLM-generated data to update the current data for the next round of prompting and generation, which is optional. 4) Training: leverage the LLM-generated data to train news recommendation models.

% Prompt design forms the foundation of GENRE, and the iterative generation and updating mechanism allows for an expansive and complex design space. In the following, we show examples of prompts designed under GENRE for news summarization, user profile modeling, and personalized news generation.

\subsection{LLMs as Content Summarizer}

Large language models are capable of summarizing text content into concise phrases or sentences, due to their training on vast amounts of natural language data and summarization tasks.
Moreover, entities like the names of individuals and locations may have appeared infrequently in the original dataset, making it challenging to learn their representations with traditional methods.
However, large language models can associate them more effectively with knowledge learned during pretraining.

By providing the content title, abstract, and category as input, the large language model produces a more informative title as output, as illustrated in~\autoref{fig:prompts}.
During downstream training, the enhanced content title will replace the original one and be used as one of the input features for the content encoder (\autoref{fig:once}).

\subsection{LLMs as User Profiler}

The user profile generally refers to their preferences and characteristics, such as age, gender, topics of interest, and geographic location.
They are usually not provided in the anonymized dataset due to privacy policies.
Large language models are capable of understanding the browsing history and analyze an outline of the user profile.

As depicted in~\autoref{fig:prompts}, the large language model produces topics and regions of interest when given the user browsing history.
In this example, GPT-3.5 infers that the user may be interested in the region of ``Florida'', based on the word ``Miami'' in the news.
While ``Miami'' may have a low occurrence in the dataset, ``Florida'' is more frequently represented and therefore more likely to be connected to other news or users for collaborative filtering.

To incorporate the inferred user profile into the recommendation model, we first fuse the topics and regions of interest into an interest vector $\mathbf{v}_i$, defined by:
\begin{gather}
    \mathbf{v}_i = \left[\mathrm{POOL}\left(\mathbf{E}_\text{topics}\right); \mathrm{POOL}\left(\mathbf{E}_\text{regions}\right)\right] \in \mathbb{R}^{2 \times d},
\end{gather}
where $\mathrm{POOL}$ is the average pooling operation, $\mathbf{E}_\text{topics}$ and $\mathbf{E}_\text{regions}$ are the embedding matrices of the interested topics and regions, and $\left[;\right]$ is the vector concatenation operation.
Then, the interest vector $\mathbf{v}_i$ will be combined with the user vector $\mathbf{v}_u$ learned from the history encoder (shown in~\autoref{fig:once}) to form the interest-aware user vector $\mathbf{v}_{iu}$ as follows:
\begin{gather}
    \mathbf{v}_{iu} = \mathrm{MLP}\left(\left[\mathbf{v}_u; \mathbf{v}_i\right]\right) \in \mathbb{R}^{d},
\end{gather}
where $\mathrm{MLP}$ is a multi-layer perceptron with ReLU activation.
Finally, the interest-aware user vector will replace the original user vector to participate in the click probability prediction.

\subsection{LLMs as Personalized Content Generator}

% The cold-start problem, which is well-known for its difficulties, occurs when new users\footnote{Following~\cite{he2016ups}, we use the term ``new users'' to refer to users with less than 5 news articles in their reading history.} have limited interaction data, making it difficult for the user encoder to capture their characteristics and ultimately weakening its ability to model warm users~\footnote{We use warm user to represent the user who has browsed more than five news articles in the history.}.
Recent studies~\cite{dai2023auggpt,ubani2023zeroshotdataaug} have shown that large language models possess exceptional capabilities to learn from few examples.
Hence, we propose to use GPT-3.5 to model the distribution of user-interested content given very limited browsing history data.
Specifically, we use it as a personalized content generator to generate synthetic content that may be of interest to new users\footnote{Following~\cite{he2016ups}, we use ``new users'' to refer to users with no more than five contents in browsing history.} have limited interaction data, making it difficult for the user encoder to capture their characteristics and ultimately weakening its ability to model warm users\footnote{We use ``warm user'' to represent the user who has browsed more than five contents.}, enhancing their historical interactions and allowing the history encoder to learn effective user representations.

% The prompt displayed in~\autoref{fig:pt-ig} serves as a guide for the personalized news generator, allowing the LLM to create synthetic news pieces tailored to the user's interests. The generated news pieces (indicated by the yellow news vectors in~\autoref{fig:ov-ig}) are incorporated into the user historical sequence, which will be encoded and fed to the user encoder to generate the user vector. 

\subsection{Chain-based Generation}

While we have shown several examples of ``one-pass generation'', it is worth noting that large language models allow iterative generation and updating.
The data generated by the large language models can be leveraged to enhance the quality of current data, which can subsequently be utilized in the next round of prompting and generation in an iterative fashion.

We design a chain-based personalized content generator by combining the one-pass user profiler and personalized content generator.
Specifically, we first use the GPT-3.5 to generate the interested topics and regions of a user, which are then combined with the user history to prompt the large language model to generate synthetic content pieces.
The user profile helps the large language models to engage in chain thinking, resulting in synthetic content that better matches the user's interests than the one-pass prompting.
\definecolor{darkgreen}{RGB}{30, 160, 30}
\newcommand{\improv}[1]{\small{\color{darkgreen}{#1}}}
\definecolor{darkblue}{RGB}{80, 80, 180}
\newcommand{\newfeat}[1]{\color{darkblue}{#1}}
\definecolor{darkred}{RGB}{180, 40, 40}
\newcommand{\bad}[1]{\color{darkred}{#1}}

\definecolor{c4}{rgb}{0.65, 0.25, 0.21}
\definecolor{c3}{rgb}{0.94, 0.76, 0.64}
\definecolor{c2}{rgb}{0.50, 0.76, 0.89}
\definecolor{c1}{rgb}{0.51, 0.72, 0.61}

\newcommand{\first}[1]{\textbf{#1}}

\section{Experiments}\label{sec:exp}

\begin{table}
\renewcommand{\arraystretch}{1.2}
\centering
\setlength\tabcolsep{2pt}
\caption{\label{tab:dataset-statistics} Data statistics. We use ``user$_n$'' to denote new users. Green numbers signify improvements over the original dataset, while blue numbers indicate the values of newly introduced features.}
% \resizebox{1.0\linewidth}{!}{
% \begin{tabular}{rlrl}
% \toprule
% \multicolumn{4}{c}{\textbf{MIND}} \\
% \midrule
% \#news & 65,238 & \#users & 94,057 \\
% \#new user & 20,110 & \#new user ratio & 0.21 \\
% \#pos & 347,727 & \#neg & 8,236,715 \\
% \#category & 18 & \#subcategory & 270 \\
% tokens/title & 13.56 & tokens/abstract & 31.86 \\
% news/user & 14.98 & news/new user & 3.19 \\
% \midrule
% \multicolumn{4}{c}{\textbf{MIND-NS}} \\
% \midrule
% tokens/title & 16.73 \improv{+3.17} &  & \\
% \midrule
% \multicolumn{4}{c}{\textbf{MIND-UP}} \\
% \midrule
% \#topics & 1,021 & \#regions & 117 \\
% topics/user & 4.82 & regions/user & 0.29 \\
% \midrule
% \multicolumn{4}{c}{\textbf{MIND-NG}} \\
% \midrule
% \#news & 105,458 \improv{+40,220} & news/new user & 5.19 \improv{+2.00} \\
% \bottomrule
% \end{tabular}
% }
\resizebox{\linewidth}{!}{
\begin{tabular}{r|cc|r|cc}
\toprule
\textbf{\textbf{Dataset}} & \textbf{\textbf{MIND}} & \textbf{\textbf{Goodreads}} & \textbf{\textbf{Dataset}} & \textbf{\textbf{MIND}} & \textbf{\textbf{Goodreads}} \\
\midrule
\multicolumn{3}{c|}{\textit{Original}} 
& \multicolumn{3}{c}{\textit{Content Summarizer (CS)}} \\

\midrule
\textbf{\# content} & 65,238 & 16,833 
& \textbf{tokens/title} & \improv{+3.17} & - \\

\textbf{tokens/title} & 13.56 & 6.10 
& \textbf{tokens/desc} & - & \newfeat{29.28} \\

\cmidrule{4-6}
\textbf{\# users} & 94,057 & 23,089 
& \multicolumn{3}{c}{\textit{User Profiler (UP)}} \\
\cmidrule{4-6}

\textbf{\# new user} & 20,110 & 2,306
& \textbf{topics/user} & \newfeat{4.82} & \newfeat{4.55} \\

\textbf{content/user} & 14.98 & 7.81  
& \textbf{regions/user} & \newfeat{0.29} & - \\

\cmidrule{4-6}
\textbf{content/user$_n$} & 3.19 & 3.03 
& \multicolumn{3}{c}{\textit{Personalized Content Generator (CG)}} \\
\cmidrule{4-6}

\textbf{\# pos} & 347,727 & 273,888
& \textbf{\#content} & \improv{+40,220} & \improv{+4,612} \\

\textbf{\# neg} & 8,236,715 & 485,233
& \textbf{content/user$_n$} & \improv{+2.00} & \improv{+2.00} \\

\bottomrule
\end{tabular}
}

\end{table}

\begin{table*}[t]
\renewcommand{\arraystretch}{1.2} 
\setlength\tabcolsep{3pt}
\caption{
%Performance comparison among the original recommenders, recommenders boosted by open-source LLM, recommenders boosted by closed-source LLM, and recommenders boosted by both types of LLM. More precisely, in the open-source LLM group, BERT$_{12L}$ method is proposed by PLM-NR~\cite{plmnr}.
%In the closed-source LLM group, we use CS, UP, NG to represent dataset enhanced by content summarizer, user profiler, and personalized content generator, respectively. We also use ALL to indicate the dataset enhanced by CS, UP, and NG. We bold the best results.
%Performance comparison of original recommenders, open-source LLM boosted recommenders, closed-source LLM boosted recommenders, and dual LLM boosted recommenders (i.e., ONCE).In the open-source LLM group, we present BERT$_{12L}$ as introduced by PLM-NR~\cite{plmnr}.For the closed-source LLM group, we use ``CS'', ``UP'', and ``CG'' to denote datasets enhanced by content summarizer, user profiler, and personalized content generator, respectively.Additionally, ``ALL'' signifies the dataset enriched by CS, UP, and CG. Best results are highlighted in bold.
Performance comparison among original recommenders, recommenders enhanced by open-source LLMs (i.e., DIRE), those enhanced by closed-source LLMs (i.e., GENRE), and those boosted by both types of LLMs (i.e., ONCE).
In the open-source LLM category, we reference the BERT$_{12L}$ approach as detailed by PLM-NR~\cite{plmnr}.
Within the closed-source LLM category, the abbreviations ``CS'', ``UP'', ``CG'', and ``UP$\rightarrow$CG'' represent datasets augmented by the one-pass content summarizer, one-pass user profiler, one-pass personalized content generator, and chain-based personalized content generator, respectively.
Furthermore, ``ALL'' denotes a dataset that incorporates enhancements from CS, and UP$\rightarrow$CG.
The top-performing results are emphasized in bold.}
\label{tab:big-table}

\resizebox{\linewidth}{!}{
\begin{tabular}{crcccccccccccccccc}
\toprule[1pt]
 & & \multicolumn{4}{c}{\textbf{NAML} (\citeyear{naml})} & \multicolumn{4}{c}{\textbf{NRMS} (\citeyear{nrms})} & \multicolumn{4}{c}{\textbf{Fastformer} (\citeyear{wu2021fastformer})} & \multicolumn{4}{c}{\textbf{MINER} (\citeyear{li2022miner})} \\
\cmidrule(lr){3-6} \cmidrule(lr){7-10} \cmidrule(lr){11-14} \cmidrule(lr){15-18} 
% \cmidrule(lr){4-7} \cmidrule(lr){8-11} \cmidrule(lr){12-15} \cmidrule(lr){16-19} 
& & \textbf{AUC} & \textbf{MRR} & \textbf{N@5} & \textbf{N@10} & \textbf{AUC} & \textbf{MRR} & \textbf{N@5} & \textbf{N@10} & \textbf{AUC} & \textbf{MRR} & \textbf{N@5} & \textbf{N@10} & \textbf{AUC} & \textbf{MRR} & \textbf{N@5} & \textbf{N@10} \\
\midrule[1pt]
\multicolumn{18}{c}{\textbf{\textit{MIND dataset}}} \\
\midrule
% \multirow{12}{*}{\textbf{MIND}}
\multicolumn{2}{r}{\textbf{Original}}
    & 61.75 & 30.60 & 31.35 & 37.85 
    & 61.71 & 30.20 & 30.98 & 37.42 
    & 62.26 & 31.14 & 31.90 & 38.32
    & 63.88 & 32.19 & 33.04 & 39.45 \\
\cmidrule{1-18}
\multicolumn{1}{l}{\multirow{3}{*}{\textbf{DIRE}}}
& \textbf{BERT}$_\text{12L}$~\cite{plmnr} 
    & 65.32 & 33.16 & 34.29 & 40.35 
    & 64.08 & 31.24 & 32.35 & 38.66
    & 65.48 & 32.47 & 33.41 & 39.75
    & 65.82 & 32.77 & 34.02 & 40.19 \\
& \textbf{LLaMA}$_\text{7B}$ \textit{(Ours)} 
    & 68.34 & 35.80 & 37.60 & 43.48 
    & 68.50 & 36.21 & 38.11 & 43.91 
    & 68.55 & 36.59 & 38.38 & 44.06
    & 68.70 & 36.58 & 38.49 & 44.18 \\
& \textbf{LLaMA}$_\text{13B}$ \textit{(Ours)} 
    & 68.23 & 35.99 & 37.93 & 43.77
    & 68.45 & 36.15 & 38.02 & 43.88
    & 68.51 & 36.37 & 38.20 & 44.02
    & 68.59 & 36.46 & 38.38 & 44.05 \\
\cmidrule{1-18}
\multicolumn{1}{l}{\multirow{5}{*}{\textbf{GENRE}}}
& \textbf{CS} \textit{(Ours)} 
    & 63.73 & 31.83 & 32.94 & 39.24
    & 63.85 & 31.57 & 32.35 & 38.80 
    & 64.73 & 32.81 & 33.68 & 40.06
    & 65.71 & 33.59 & 34.90 & 40.96 \\
& \textbf{UP} \textit{(Ours)} 
    & 62.19 & 30.90 & 31.78 & 38.26 
    & 61.90 & 30.60 & 31.54 & 37.66 
    & 63.40 & 31.94 & 32.76 & 39.15
    & 64.45 & 32.09 & 33.14 & 39.54\\
& \textbf{CG} \textit{(Ours)} 
    & 62.93 & 30.83 & 32.10 & 38.34 
    & 63.04 & 31.00 & 31.84 & 38.22 
    & 64.69 & 32.28 & 33.31 & 39.76
    & 64.21 & 32.30 & 33.57 & 39.91 \\
& \textbf{UP}$\rightarrow$\textbf{CG} \textit{(Ours)} 
    & 63.61 & 31.58 & 32.63 & 39.07
    & 62.95 & 32.00 & 32.80 & 39.00
    & 64.82 & 32.44 & 33.51 & 39.93
    & 64.73 & 33.09 & 34.10 & 40.32
\\
& \textbf{ALL} \textit{(Ours)} 
    & 63.88 & 32.17 & 33.14 & 39.37
    & 63.71 & 32.14 & 33.11 & 39.43 
    & 66.70 & 34.20 & 35.81 & 41.78
    & 66.46 & 34.20 & 35.47 & 41.48 \\
\cmidrule{1-18}
% & \textbf{UP$\rightarrow$NG} 
%     & 63.61 & 31.58 & 32.63 & 39.07
%     & 62.95 & 32.00 & 32.80 & 39.00
%     & - & - & - & - \\
% & \textbf{ALL} 
%     & \textbf{63.88} & \textbf{32.17} & \textbf{33.14} & \textbf{39.37}
%     & 63.71 & \textbf{32.14} & \textbf{33.11} & \textbf{39.43} 
%     & \textbf{66.70} & \textbf{34.20} & \textbf{35.81} & \textbf{41.78} \\
\multicolumn{2}{r}{\textbf{ONCE} \textit{(ours)}}
    & \first{68.62} & \first{36.50} & \first{38.31} & \first{44.05}
    & \first{68.74} & \first{36.66} & \first{38.60} & \first{44.37}
    & \first{68.83} & \first{36.68} & \first{38.56} & \first{44.35}
    & \first{68.92} & \first{36.74} & \first{38.72} & \first{44.48} \\
\multicolumn{2}{r}{\textbf{Improvement (\%) over Original}}
    & \first{11.13\%} & \first{19.28\%} & \first{22.20\%} & \first{16.38\%}
    & \first{11.39\%} & \first{21.39\%} & \first{24.60\%} & \first{18.57\%}
    & \first{10.55\%} & \first{17.79\%} & \first{20.88\%} & \first{15.74\%}
    & \first{7.89\%} & \first{14.13\%} & \first{17.19\%} & \first{12.75\%} \\
\multicolumn{2}{r}{\textbf{Improvement (\%) over BERT$_{12L}$}}
    & \first{5.05\%} & \first{10.07\%} & \first{11.72\%} & \first{9.17\%} 
    & \first{7.27\%} & \first{17.35\%} & \first{19.32\%} & \first{14.77\%} 
    & \first{5.12\%} & \first{12.97\%} & \first{15.41\%} & \first{11.57\%}
    & \first{4.71\%} & \first{12.11\%} & \first{13.82\%} & \first{10.67\%} \\
\midrule[1pt] 
\multicolumn{18}{c}{\textbf{\textit{Goodreads dataset}}} \\
\midrule
% \multirow{12}{*}{\textbf{Goodreads}}
\multicolumn{2}{r}{\textbf{Original}}
    & 66.47 & 75.75 & 58.49 & 82.20
    & 68.95 & 77.05 & 60.62 & 83.16 
    & 70.85 & 78.37 & 62.90 & 84.15
    & 71.03 & 78.46 & 63.09 & 84.20 \\
\cmidrule{1-18}

\multicolumn{1}{l}{\multirow{3}{*}{\textbf{DIRE}}}
& \textbf{BERT}$_\text{12L}$~\cite{plmnr} 
    & 70.68 & 78.17 & 62.26 & 83.99
    & 71.80 & 78.87 & 63.62 & 84.51
    & 72.47 & 79.29 & 64.45 & 84.82
    & 73.36 & 80.08 & 65.19 & 85.25 \\
& \textbf{LLaMA}$_\text{7B}$ \textit{(Ours)} 
    & 77.01 & 82.74 & 71.09 & 89.39 
    & 75.90 & 81.75 & 69.13 & 86.65
    & 76.52 & 82.31 & 70.48 & 87.03
    & 76.45 & 82.46 & 70.31 & 86.92 \\
& \textbf{LLaMA}$_\text{13B} $\textit{(Ours)} 
    & 77.43 & 83.05 & 71.56 & 87.61
    & 77.57 & 82.96 & 71.41 & 87.55
    & 77.46 & 83.00 & 71.36 & 87.58
    & 77.50 & 83.07 & 71.44 & 87.64 \\

\cmidrule{1-18}

\multicolumn{1}{l}{\multirow{5}{*}{\textbf{GENRE}}}
& \textbf{CS} \textit{(Ours)} 
    & 67.68 & 76.41 & 59.64 & 82.69
    & 69.77 & 77.57 & 61.54 & 83.35  
    & 71.41 & 78.77 & 63.70 & 84.43
    & 71.96 & 79.09 & 64.30 & 84.72 \\
& \textbf{UP} \textit{(Ours)} 
    & 68.45 & 76.91 & 60.70 & 83.08 
    & 69.45 & 77.58 & 61.89 & 83.57 
    & 71.15 & 78.68 & 63.86 & 84.39
    & 71.67 & 78.85 & 63.94 & 84.50 \\
& \textbf{CG} \textit{(Ours)} 
    % & 67.77 & 76.50 & 59.78 & 82.77 
    % & 72.26 & 79.27 & 64.72 & 84.82 
    % & 73.51 & 80.22 & 66.59 & 85.52 
    % & 72.10 & 79.16 & 64.36 & 84.71 \\
    & 66.94 & 76.10 & 59.26 & 82.47
    & 70.09 & 77.95 & 62.34 & 83.83 
    & 71.08 & 78.53 & 63.38 & 84.26
    & 71.81 & 78.89 & 63.99 & 84.53 \\
& \textbf{UP}$\rightarrow$\textbf{CG} \textit{(Ours)} 
    & 67.98 & 76.78 & 60.56 & 82.96 
    & 69.95 & 77.79 & 62.07 & 83.71
    & 71.88 & 79.02 & 64.10 & 84.63
    & 71.79 & 78.93 & 63.97 & 84.56 \\
& \textbf{ALL} \textit{(Ours)} 
    & 68.95 & 77.25 & 61.19 & 83.32
    & 72.07 & 79.13 & 64.46 & 84.72
    & 73.23 & 79.97 & 66.07 & 85.33
    & 73.21 & 79.91 & 65.73 & 85.29 \\
% & \textbf{ALL} \textit{(Ours)} 
%     & 69.22 & 77.39 & 61.36 & 83.35
%     & 73.03 & 79.71 & 65.69 & 85.10
%     & 74.12 & 80.63 & 67.03 & 85.79
%     & 73.19 & 79.64 & 65.58 & 85.04 \\
    
\cmidrule{1-18}

\multicolumn{2}{r}{\textbf{ONCE} \textit{(ours)}}
    & \first{77.63} & \first{83.13} & \first{71.65} & \first{87.66}
    & \first{77.89} & \first{83.31} & \first{71.89} & \first{87.79}
    & \first{78.03} & \first{83.52} & \first{72.52} & \first{87.96}
    & \first{77.82} & \first{83.35} & \first{71.96} & \first{87.85} \\
\multicolumn{2}{r}{\textbf{Improvement (\%) over Original}}
    & \first{16.79\%} & \first{9.74\%} & \first{22.50\%} & \first{6.64\%} 
    & \first{12.97\%} & \first{8.12\%} & \first{18.59\%} & \first{5.57\%} 
    & \first{10.13\%} & \first{6.57\%} & \first{15.29\%} & \first{4.53\%}
    & \first{9.56\%} & \first{6.23\%} & \first{14.06\%} & \first{4.33\%} \\
\multicolumn{2}{r}{\textbf{Improvement (\%) over BERT$_{12L}$}}
    & \first{9.83\%} & \first{6.35\%} & \first{15.08\%} & \first{4.37\%} 
    & \first{8.48\%} & \first{5.63\%} & \first{13.00\%} & \first{3.88\%} 
    & \first{7.67\%} & \first{5.33\%} & \first{12.52\%} & \first{3.70\%}
    & \first{6.08\%} & \first{4.08\%} & \first{10.39\%} & \first{3.05\%} \\
\bottomrule[1pt]
\end{tabular}
}
\end{table*}

\subsection{Experimental Setup}

\textbf{Datasets.}
We conduct experiments on two real-world content-based recommendation dataset, i.e., news recommendation dataset MIND~\cite{mind} and book recommendation dataset Goodreads~\cite{goodreads}.
In~\autoref{tab:dataset-statistics}, we present the statistics of both the original dataset and the augmented versions.
We use LLaMA-7B and LLaMA-13B models~\cite{llama} as our open-source large language models, and GPT-3.5\footnote{~\url{https://platform.openai.com/docs/guides/chat}} as our closed-source model.
For the augmented datasets, only the attributes that are different than the original datasets are shown in \autoref{tab:dataset-statistics}.
For the Goodreads dataset, the content summarizer is used for the book description generation, given only the book title.

\begin{figure}[h]
    \centering
    \setlength\tabcolsep{0pt}
    \resizebox{.9\linewidth}{!}{
    \begin{tabular}{m{0.25\textwidth}m{0.25\textwidth}}
    \multicolumn{2}{c}{
        \resizebox{1.0\linewidth}{!}{
            \begin{tikzpicture}
    \begin{customlegend}[
        legend columns=4,
        legend style={
            align=left,
            draw=none,
            column sep=2ex
        },
        legend entries={
            \textsc{\small{ONCE}},
            \textsc{\small{LLaMA$_\text{13B}$}},
            \textsc{\small{LLaMA$_\text{7B}$}},
            \textsc{\small{BERT$_\text{12L}$}},
        }]
        \addlegendimage{c4,mark=x,solid,line legend}
        \addlegendimage{c3,mark=x,solid,line legend}
        \addlegendimage{c2,mark=x,solid,line legend}
        \addlegendimage{c1,mark=x,solid,line legend}
    \end{customlegend}
\end{tikzpicture}
        }
    } \\
    \begin{subfigure}{0.25\textwidth}
        \resizebox{1.0\linewidth}{!}{
            % \begin{figure}
% \centering
% \resizebox{0.75\linewidth}{!}{
\begin{tikzpicture}
\begin{axis}[    
    xlabel={Epoch},
    ylabel={AUC},    
    ymin=70,    
    ymax=83,    
    grid=both,
    grid style=dashed,
    legend pos=south east, 
]

    \draw[
            c2!80!black,
            dashed,
            line width=1mm,
            opacity=0.7
        ] (0,177) -- (100,177);

    \addplot[c4,line width=2pt] coordinates {
        (1, 74.92)
        (2, 77.74)
        (3, 79.24)
        (4, 80.09)
        (5, 80.45)
        (6, 80.90)
        (7, 81.32)
        (8, 81.02)
        (9, 81.39)
        (10, 81.50)
        (11, 81.79)
        (12, 81.94)
        (13, 82.03)
    };
    % \addlegendentry{ONCE}

     \addplot[c2,line width=2pt] coordinates {
        (1, 63.12)
        (2, 66.30)
        (3, 68.92)
        (4, 70.35)
        (5, 71.67)
        (6, 72.13)
        (7, 72.68)
        (8, 72.85)
        (9, 73.44)
        (10, 73.69)
        (11, 73.82)
        (12, 74.39)
        (13, 74.71)
        (14, 75.26)
        (15, 75.78)
    };
    
    \addplot[c3,line width=2pt] coordinates {
        (1, 72.69)
        (2, 75.36)
        (3, 77.46)
        (4, 78.75)
        (5, 79.16)
        (6, 80.47)
        (7, 80.55)
        (8, 80.75)
        (9, 80.71)
    };
    % \addlegendentry{LLaMA$_\text{13B}$}
    
    \addplot[c1,line width=2pt] coordinates {
        (1, 63.41)
        (2, 65.73)
        (3, 67.46)
        (4, 68.58)
        (5, 69.57)
        (6, 69.77)
        (7, 70.52)
        (8, 71.00)
        (9, 71.88)
        (10, 72.01)
        (11, 72.39)
        (12, 72.99)
        (13, 73.17)
        (14, 73.40)
        (15, 73.74)
        % (16, 74.11)
        % (17, 74.41)
        % (18, 75.09)
        % (19, 75.29)
        % (20, 75.32)
        % (21, 75.30)
    };
    % \addlegendentry{BERT$_\text{12L}$}

  \end{axis}
\end{tikzpicture}
% }
% \caption{\label{fig:epoch}Influence of the number of generated news articles on the AUC metric over four base models.}
% \end{figure}
        }
        \caption{\label{fig:epoch-nrms}NRMS}
    \end{subfigure} & 
    \begin{subfigure}{0.25\textwidth}
        \resizebox{1.0\linewidth}{!}{
            % \begin{figure}
% \centering
% \resizebox{0.75\linewidth}{!}{
\begin{tikzpicture}
\begin{axis}[    
    xlabel={Epoch},
    ylabel={AUC},    
    ymin=70,    
    ymax=83,    
    grid=both,
    grid style=dashed,
    legend pos=south east, 
]

    \draw[
            c2!80!black,
            dashed,
            line width=1mm,
            opacity=0.7
        ] (0,175) -- (150,175);

    \addplot[c4,line width=2pt] coordinates {
        (1, 71.49)
        (2, 75.65)
        (3, 77.11)
        (4, 77.25)
        (5, 78.96)
        (6, 79.62)
        (7, 79.35)
        (8, 80.29)
        (9, 80.50)
        (10, 80.86)
        (11, 81.15)
        (12, 81.34)
        (13, 81.57)
        (14, 81.79)
        (15, 81.63)
    };
    % \addlegendentry{ONCE}
    
    \addplot[c3,line width=2pt] coordinates {
        (1, 70.27)
        (2, 72.34)
        (3, 74.67)
        (4, 75.44)
        (5, 76.65)
        (6, 77.44)
        (7, 77.96)
        (8, 78.16)
        (9, 79.09)
        (10, 79.21)
        (11, 79.70)
        (12, 79.95)
        (13, 80.09)
        (14, 80.07)
        (15, 80.46)
        (15, 80.27)
    };
    % \addlegendentry{LLaMA$_\text{13B}$}
    
    % \addplot[c3,line width=2pt] coordinates {
    %     (1, 72.69)
    %     (2, 75.36)
    %     (3, 77.46)
    %     (4, 78.75)
    %     (5, 79.16)
    %     (6, 80.47)
    %     (7, 80.55)
    %     (8, 80.75)
    %     (9, 80.71)
    % };
    % \addlegendentry{LLaMA$_\text{13B}$}
    
    \addplot[c1,line width=2pt] coordinates {
        (1, 63.00)
        (2, 65.49)
        (3, 67.45)
        (4, 68.62)
        (5, 69.14)
        (6, 70.09)
        (7, 70.58)
        (8, 71.50)
        (9, 71.68)
        (10, 72.09)
        (11, 72.60)
        (12, 72.81)
        (13, 72.74)
        (14, 73.77)
        (15, 73.69)
    };
    % \addlegendentry{BERT$_\text{12L}$}

   \addplot[c2,line width=2pt] coordinates {
        (1, 63.29)
        (2, 66.58)
        (3, 69.92)
        (4, 70.95)
        (5, 72.18)
        (6, 73.39)
        (7, 73.69)
        (8, 74.47)
        (9, 74.79)
        (10, 74.81)
        (11, 75.43)
        (12, 75.74)
        (13, 76.18)
        (14, 76.48)
        (15, 77.04)
        % (16, 77.24)
        % (17, 77.40)
        % (18, 77.92)
        % (19, 78.08)
        % (20, 78.24)
        % (21, 78.29)
        % (22, 78.72)
        % (23, 79.13)
        % (24, 79.47)
    };
    % \addlegendentry{LLaMA$_\text{7B}$}
    
  \end{axis}
\end{tikzpicture}
% }
% \caption{\label{fig:epoch}Influence of the number of generated news articles on the AUC metric over four base models.}
% \end{figure}
        }
        \caption{\label{fig:epoch-fastformer}Fastformer}
    \end{subfigure}
    \end{tabular}
    }
    
    \caption{\label{fig:epoch} Training curves for open-source LLMs and ONCE. The y-axis AUC value is evaulated on the validation set.}
\end{figure}

\textbf{Recommendation Models.} We evaluate the effectiveness of proposed ONCE method with three popular content-based recommendation models, namely NAML~\cite{naml}, NRMS~\cite{nrms}, and Fastformer~\cite{wu2021fastformer}.
We also compare with PLM-NR~\cite{plmnr} method, which replaces the original content encoder with small-scale pretrained language models such as BERT~\cite{bert}.

\textbf{Evaluation Metrics.}
We follow the common practice~\cite{liu2022prec,nrms,plmnr} to evaluate the effectiveness of news recommendation models with the widely used metrics, i.e., AUC~\cite{auc}, MRR~\cite{mrr} and nDCG~\cite{ndcg}.
In this work, we use nDCG@1 and nDCG@5 for evaluation on the Goodreads dataset, and nDCG@5 and nDCG@10 for evaluation on the MIND dataset shortly denoted as N@1, N@5 and N@10, respectively.

% \textbf{News Features.} To incorporate image information into text-based news recommendation models, we use a pretrained image encoder~\cite{clip} to extract image features, which we treat as a news-specific token.
% We also treat the category and subcategory as special tokens that do not undergo tokenization.
% Then, we concatenate these features (i.e., title, image, and category) to form the input sequence for the news encoder.
% For the NAML model, since its original news encoder already incorporates the category information, we only concatenate the image and title features.

\begin{table*}[]
\centering
\renewcommand\arraystretch{1.2}
\setlength\tabcolsep{4pt}

\caption{Influence of the number of frozen layers on three open-source LLMs. Best results are highlighted in bold, while results inferior to the respective base models are indicated in \textcolor{darkred}{red}. ``F/T'' denotes the number of frozen and tuning layers, respectively.}\label{tab:layer}

\resizebox{\linewidth}{!}{
\begin{tabular}{clcccccccccccccccc}
\toprule
& & \multicolumn{4}{c}{\textbf{NAML} (\citeyear{naml})} & \multicolumn{4}{c}{\textbf{NRMS} (\citeyear{nrms})} & \multicolumn{4}{c}{\textbf{Fastformer} (\citeyear{wu2021fastformer})} & \multicolumn{4}{c}{\textbf{MINER} (\citeyear{li2022miner})} \\
% \cmidrule(lr){4-7} \cmidrule(lr){8-11} \cmidrule(lr){12-15} \cmidrule(lr){16-19}
\cmidrule(lr){3-6} \cmidrule(lr){7-10} \cmidrule(lr){11-14} \cmidrule(lr){15-18}
\textbf{Encoder} & \textbf{F/T} & \textbf{AUC} & \textbf{MRR} & \textbf{N@5} & \textbf{N@10} & \textbf{AUC} & \textbf{MRR} & \textbf{N@5} & \textbf{N@10} & \textbf{AUC} & \textbf{MRR} & \textbf{N@5} & \textbf{N@10} & \textbf{AUC} & \textbf{MRR} & \textbf{N@5} & \textbf{N@10} \\
\midrule[1pt]
\multicolumn{18}{c}{\textbf{\textit{MIND dataset}}} \\
\midrule
% \multirow{1}{*}{\textbf{MIND}} &
\textbf{Original} & -
    & 61.75 & 30.60 & 31.35 & 37.85 
    & 61.71 & 30.20 & 30.98 & 37.42 
    & 62.26 & 31.14 & 31.90 & 38.32 
    & 63.88 & 32.19 & 33.04 & 39.45 \\
\cmidrule{1-18}
\multirow{3}{*}{\textbf{BERT}$_{12L}$}
 & 12/0 
    & 65.32 & 33.16 & 34.29 & 40.35 
    & 64.08 & 31.24 & 32.35 & 38.66 
    & 64.25 & 32.05 & 32.88 & 39.17
    & 64.75 & 32.44 & 33.60 & 39.87 \\
& 11/1
    & 65.10 & 32.86 & 33.99 & 40.19
    & 62.59 & 31.46 & 32.09 & 38.61 
    & 65.48 & 32.47 & 33.41 & 39.75
    & 65.82 & 32.77 & 34.02 & 40.19 \\
& 10/2
    & 63.79 & 32.27 & 32.95 & 39.40 
    & 62.68 & 30.95 & 31.61 & 37.89
    & 63.41 & 31.57 & 32.56 & 38.92
    & 64.01 & 31.69 & 32.82 & 39.17 \\
 % & 8/4  &  &  &  &  & \\
\cmidrule{1-18}
\multirow{3}{*}{\textbf{LLaMA}$_{7B}$}
 % & 32/0* & 65.53 & 33.17 & 34.65 & 40.87 & \\
 & 32/0
    & 67.78 & 35.17 & 36.84 & 42.78 
    & 68.10 & 35.33 & 36.91 & 43.04
    & 67.83 & 35.19 & 36.57 & 42.59
    & 67.96 & 35.28 & 36.72 & 42.80 \\
& 31/1
    & \first{68.34} & 35.80 & 37.60 & 43.48 
    & 68.33 & 35.81 & 37.43 & 43.37
    & 68.51 & 36.56 & \textbf{38.46} & \textbf{44.15}
    & 68.45 & 36.41 & 38.25 & 43.93 \\
& 30/2
    & 68.18 & \first{36.09} & 37.76 & 43.65
    & \first{68.50} & \first{36.21} & \first{38.11} & \first{43.91}
    & 68.55 & \first{36.59} & 38.38 & 44.06
    & \first{68.70} & \first{36.58} & \first{38.49} & \first{44.18} \\
\cmidrule{1-18}
\multirow{3}{*}{\textbf{LLaMA}$_{13B}$}
 % & 40/0* &  &  &  &  & \\
 & 40/0 
    & 68.23 & 35.99 & \first{37.93} & \first{43.77}
    & 68.45 & 36.15 & 38.02 & 43.88
    & 68.51 & 36.37 & 38.20 & 44.02
    & 68.59 & 36.46 & 38.38 & 44.05 \\
& 39/1
    & 67.66 & 35.73 & 37.59 & 43.35
    & 68.23 & 36.05 & 37.97 & 43.72
    & \textbf{68.60} & 36.45 & 38.27 & 43.96
    & 68.53 & 36.37 & 38.21 & 44.00 \\
& 38/2
    & 68.19 & 36.07 & 37.89 & 43.68
    & 68.30 & 36.13 & 37.95 & 43.74
    & 68.19 & 35.96 & 37.72 & 43.50
    & 67.83 & 35.88 & 37.64 & 43.45 \\
 % & 36/4  &  &  &  &  & \\
\midrule[1pt]
\multicolumn{18}{c}{\textbf{\textit{Goodreads dataset}}} \\
\midrule
% \multirow{11}{*}{\textbf{Goodreads}} &
\textbf{Original} & -
    & 66.47 & 75.75 & 58.49 & 82.20
    & 68.95 & 77.05 & 60.62 & 83.16 
    & 70.85 & 78.37 & 62.90 & 84.15
    & 71.03 & 78.46 & 63.09 & 84.20 \\
\cmidrule{1-18}
\multirow{4}{*}{\textbf{BERT}$_{12L}$}
 & \bad{12/0} 
    & \bad{62.05} & \bad{72.82} & \bad{53.37} & \bad{80.06} 
    & \bad{64.49} & \bad{74.38} & \bad{56.47} & \bad{81.22} 
    & \bad{66.83} & \bad{75.85} & \bad{58.68} & \bad{82.35}
    & \bad{67.11} & \bad{76.09} & \bad{58.88} & \bad{82.48} \\
& \bad{11/1}
    & \bad{62.32} & \bad{73.03} & \bad{53.82} & \bad{80.22} 
    & \bad{65.94} & \bad{75.35} & \bad{58.12} & \bad{81.94} 
    & \bad{66.23} & \bad{75.53} & \bad{58.12} & \bad{82.05}
    & \bad{66.72} & \bad{75.93} & \bad{58.60} & \bad{82.28} \\
& \bad{10/2}
    & \bad{65.22} & \bad{74.90} & \bad{57.07} & \bad{81.58} 
    & \bad{63.77} & \bad{73.94} & \bad{55.53} & \bad{80.88} 
    & \bad{67.66} & \bad{76.49} & \bad{60.03} & \bad{82.76}
    & \bad{67.94} & \bad{76.72} & \bad{60.22} & \bad{82.89} \\
& 0/12
    & 70.68 & 78.17 & 62.26 & 83.99
    & 71.80 & 78.87 & 63.62 & 84.51
    & 72.47 & 79.29 & 64.45 & 84.82
    & 73.36 & 80.08 & 65.19 & 85.25 \\
 % & 8/4  &  &  &  &  & \\
\cmidrule{1-18}
\multirow{3}{*}{\textbf{LLaMA}$_{7B}$}
 & 32/0
    & 69.29 & 77.32 & 60.89 & 83.37 
    & 71.96 & 79.19 & 64.73 & 84.77 
    & 72.25 & 79.16 & 64.35 & 84.73 
    & 71.14 & 78.53 & 63.44 & 84.27 \\
& 31/1
    & 73.82 & 80.34 & 66.51 & 85.61 
    & 75.18 & 81.23 & 68.04 & 86.27 
    & 75.80 & 81.70 & 68.69 & 86.58
    & 75.33 & 81.35 & 68.26 & 86.33 \\
& 30/2
    & 77.01 & 82.74 & 71.09 & \first{89.39} 
    & 75.90 & 81.75 & 69.13 & 86.65 
    & 76.52 & 82.31 & 70.48 & 87.03
    & 76.45 & 82.46 & 70.31 & 86.92 \\
 % & 28/4  &  &  &  &  & \\
\cmidrule{1-18}
\multirow{3}{*}{\textbf{LLaMA}$_{13B}$}
 & 40/0  
    & 70.31 & 78.00 & 62.36 & 83.88
    & 72.82 & 79.79 & 65.79 & 85.21
    & 71.66 & 78.81 & 63.52 & 84.48
    & 73.28 & 80.13 & 66.09 & 85.23 \\
& 39/1  
    & \first{77.43} & \first{83.05} & \first{71.56} & 87.61 
    & 76.55 & 82.32 & 70.08 & 87.07 
    & 76.42 & 82.36 & 70.51 & 87.10
    & 77.18 & 82.60 & 71.17 & 87.43 \\
& 38/2  
    & 76.25 & 82.18 & 69.98 & 86.98 
    & \first{77.57} & \first{82.96} & \first{71.41} & \first{87.55}
    & \first{77.46} & \first{83.00} & \first{71.36} & \first{87.58}
    & \first{77.50} & \first{83.07} & \first{71.44} & \first{87.64} \\
\bottomrule
\end{tabular}
}
\end{table*}

% \begin{table*}[]
% \renewcommand{\arraystretch}{1.2} 
% \setlength\tabcolsep{3pt}
% \caption{AUC improvement per model size (\%).}
% \begin{tabular}{crcccccccc}
% \toprule[1pt]
% & & \multicolumn{4}{c}{\textbf{\textit{MIND dataset}}} & \multicolumn{4}{c}{\textbf{\textit{Goodreads dataset}}} \\
% \cmidrule(lr){3-6} \cmidrule(lr){7-10} 
% & & \textbf{NAML} & \textbf{NRMS} & \textbf{Fastformer} & \textbf{MINER} & \textbf{NAML} & \textbf{NRMS} & \textbf{Fastformer} & \textbf{MINER} \\
% \midrule
%  & \textbf{Original}  &  &  &  &  \\
% \midrule
% \multicolumn{1}{l}{\multirow{3}{*}{\textbf{DIRE}}}
%  & \textbf{BERT}$_\text{12L}$ &  &  &  &  \\
%  & \textbf{LLaMA}$_\text{7B}$ &  &  &  &  \\
%  & \textbf{LLaMA}$_\text{13B}$ &  &  &  &  \\
% \midrule
% \multicolumn{1}{l}{\multirow{5}{*}{\textbf{GENRE}}}
% & \textbf{CS} &  &  &  & \\
% & \textbf{UP} &  &  &  & \\
% & \textbf{CG} &  &  &  & \\
% & \textbf{UP}$\rightarrow$\textbf{CG} &  &  &  & \\
% & \textbf{ALL} &  &  &  & \\
% \midrule
% & \textbf{ONCE} & & & & \\
% \bottomrule[1pt]
% \end{tabular}
% \end{table*}

\textbf{Implementation Details.}
%We utilize pretrained ``clip-vit-base-patch32'' models~\cite{clip} to extract cover image features and BertTokenizer provided by the transformers package~\cite{transformers} to tokenize textual features of news articles.
During training, we employ Adam~\cite{kingma2014adam} optimizer with a learning rate of 1e-3 for the MIND dataset and 1e-4 for the Goodreads dataset.
If the large language models are not tuned with LoRA~\cite{hu2021lora}, their learning rates are set to 1e-5.
For all models, the embedding dimension of non-LLM modules is set to 64, and the negative sampling ratio is set to 4.
We tune the hyperparameters of all base models to attain optimal performance.
We average the results of five independent runs for each model and observe the p-value smaller than 0.01.
All LLaMA-based experiments are conducted on a single NVIDIA A100 device with 80GB memory, and others on a single NVIDIA GeForce RTX 3090 device.
We release all our code and datasets\footnote{\url{https://github.com/Jyonn/ONCE}} for other researches to reproduce our work.

\subsection{Performance Comparison}

\autoref{tab:big-table} provides an overview of the performance enhancements observed across four base models on two datasets, boosted by open-source LLM, closed-source LLM, and dual LLM (i.e., ONCE) approaches.
Drawing from the results, we can derive the following observations:

\textbf{Firstly}, the open-source LLM group exhibits substantial improvements in the base models.
The pretraining of LLaMA endows it with robust semantic understanding and a wealth of content-level knowledge, including elements like book titles and geographic locations.
Additionally, its high-dimensional representation space ensures efficient encoding of extensive information within hidden states.
\textbf{Secondly}, the closed-source LLM group also demonstrates impressive performance, highlighting the efficacy of data enrichment in introducing enhanced semantic features to the dataset.
The fusion of diverse prompt techniques (i.e., ``ALL'') further amplifies model effectiveness.
\textbf{Thirdly}, our dual LLM-based ONCE method showcases additional performance gains compared to employing a single LLM, albeit the improvement is relatively modest compared to the open-source finetuning.
GPT-3.5 offers LLaMA with supplementary semantic insights, elevating its content comprehension capabilities.
However, the closed-source LLM contributes token-level discrete features, which bear less influence when juxtaposed with the continuous embedding-level representations delivered by open-source LLMs.

In addition,\autoref{fig:epoch} presents the training curves for open-source LLMs and ONCE.
Notably, ONCE (using LLaMA-13B as the backbone), leveraging closed-source LLM information, demonstrates both a stronger initial performance and quicker training efficiency.
Specifically, on the NRMS model, ONCE reaches performance equivalent to LLaMA-13B's 8th epoch by its 6th epoch, a substantial 25\% improvement.
On the Fastformer model, ONCE surpasses LLaMA-13B's 15th epoch performance by its 9th epoch, showcasing an impressive 40\% enhancement.

\subsection{Ablation Study on Open-source LLMs}

Here, we study the impact of the finetuning layers and low-rank adaptation (LoRA) on the performance of open-source LLMs.

~\autoref{tab:layer} presents a comparison of finetuning effects on the top $0\sim2$ layers of transformers across different open-source LLMs. Key findings from the results include:

\textbf{Firstly}, In most instances, substantial enhancements in recommendation models are evident even without finetuning (T=0) the LLMs.
Notably, the BERT model on the Goodreads dataset is an exception due to the unique challenge posed by book titles as content, which lacks the enriched knowledge of LLaMA, resulting in less effective representations primarily focused on literal meanings.
\textbf{Secondly}, within the MIND dataset, LLaMA-7B generally outperforms LLaMA-13B with finetuning $1\sim2$ layers.
This might stem from the relative difficulty in fine-tuning LLaMA-13B, while the 7B model sufficiently captures the semantic richness of news headlines.
Conversely, for the Goodreads dataset, LLaMA-13B demonstrates the most promising outcomes.
\textbf{Thirdly}, overall, a greater number of tuned layers correlates with improved performance, though this also entails increased training costs.

~\autoref{tab:lora} presents the influence of LoRA during the finetuning process of open-source LLMs.
Our findings indicate that, for the MIND dataset, LoRA leads to improved performance, while a different pattern emerges for the Goodreads dataset.
This divergence might be attributed to differences in the nature of the input textual data.
Goodreads employs book titles with relatively limited informative content, whereas MIND's input news headlines inherently encapsulate the core essence of the content.
Constructing a robust representation from book titles requires more nuanced adjustments of the network parameters.

\begin{table}[]
\centering
\renewcommand\arraystretch{1.2}
\setlength\tabcolsep{4pt}

\caption{Influence of the use of low-rank adaption (LoRA). The experiments are conducted over the NAML model.}\label{tab:lora}

\resizebox{\linewidth}{!}{
\begin{tabular}{ccccccccc}
\toprule
 & \multicolumn{4}{c}{\textbf{LoRa}} & \multicolumn{4}{c}{\textbf{w/o LoRa}} \\
\cmidrule(lr){2-5} \cmidrule(lr){6-9}
\textbf{Encoder} & \textbf{AUC} & \textbf{MRR} & \textbf{N@5} & \textbf{N@10} & \textbf{AUC} & \textbf{MRR} & \textbf{N@5} & \textbf{N@10} \\
\midrule[1pt]
\multicolumn{9}{c}{\textbf{\textit{MIND dataset}}} \\
\midrule
\textbf{BERT}$_\text{12l}$
 & 65.10 & 32.86 & 33.99 & 40.19 & 62.94 & 31.32 & 32.20 & 38.52 \\
\textbf{LLaMA}$_\text{7B}$
 & 68.34 & 35.80 & 37.60 & 43.48 & 67.25 & 34.28 & 36.00 & 42.12 \\
\midrule[1pt]
\multicolumn{9}{c}{\textbf{\textit{Goodreads dataset}}} \\
\midrule
\textbf{BERT}$_\text{12L}$
 & 63.18 & 76.80 & 55.37 & 80.69 & 70.68 & 78.17 & 62.26 & 83.99 \\
\textbf{LLaMA}$_\text{7B}$
 & 75.00 & 81.23 & 68.44 & 86.29 & 77.01 & 82.74 & 71.09 & 89.39 \\
\bottomrule
\end{tabular}
}
\end{table}

\begin{table}[t]
\centering
\renewcommand{\arraystretch}{1.2}
\setlength\tabcolsep{2pt}
\caption{\label{tab:ig}Effectiveness of the personalized content generator ($\text{CG}$) for both new user and warm user groups, assessed on the MIND dataset. $\text{ORI}$: training with the original data. $\text{Imp.}$: denotes the improvement realized through the personalized content generator.}

\resizebox{\linewidth}{!}{
\begin{tabular}{cccccccccc}
 \toprule
 &  & \multicolumn{4}{c}{\textbf{New User}} & \multicolumn{4}{c}{\textbf{Warm User}} \\
\cmidrule(lr){3-6} \cmidrule(lr){7-10}
 &  & \textbf{AUC} & \textbf{MRR} & \textbf{N@5} & \textbf{N@10} & \textbf{AUC} & \textbf{MRR} & \textbf{N@5} & \textbf{N@10} \\
\midrule
\multirow{3}{*}{\textbf{NAML}}
 & \textbf{ORI} & 59.24 & \textbf{32.82} & 34.24 & \textbf{40.34} & 62.21 & 30.20 & 30.83 & 37.40 \\
 & \textbf{CG} & \textbf{60.21} & 32.69 & \textbf{34.67} & 40.33 & \textbf{63.43} & \textbf{30.49} & \textbf{31.64} & \textbf{37.98} \\
 & \textbf{Imp.} & 0.97 & - & 0.43 & - & 1.22 & 0.29 & 0.81 & 0.58 \\
\midrule
\multirow{3}{*}{\textbf{NRMS}}
 & \textbf{ORI} & 59.49 & 32.75 & 33.99 & 40.09 & 62.12 & 29.74 & 30.43 & 36.93  \\
 & \textbf{CG} & \textbf{59.88} & \textbf{32.90} & \textbf{34.42} & \textbf{40.16} & \textbf{63.61} & \textbf{30.65} & \textbf{31.37} & \textbf{37.87} \\
 & \textbf{Imp.} & 0.39 & 0.25 & 0.43 & 0.07 & 1.49 & 0.91 & 0.94 & 0.94 \\
\bottomrule
\end{tabular}
}
\end{table}

\subsection{Ablation Study on Closed-source LLMs}

Here, we investigate the impact of the synthetic content data on two user groups, i.e., new user group and warm user group. From the results in~\autoref{tab:ig}, it can be seen that the personalized content generator improves the performance of both the new and warm user groups in most cases. This is because the history encoder struggles to capture the interests of new users due to their limited history, which also affects its ability to model warm users. With the generated content pieces added to the history of new users, the history encoder can better capture their interests, leading to a performance improvement on both groups.

\section{Related Works} 
\subsection{LLMs for Recommendation}

The recent advancement of Large Language Models~(LLMs) like ChatGPT and LLaMa~\cite{llama}, has triggered a new wave of interest, resulting in the development of diverse applications across multiple domains~\cite{dai2023auggpt,qureshi2023exploring,wu2023bloomberggpt}. Using self-supervised learning on large datasets, these models excel in text representation and, with transfer techniques such as fine-tuning and prompt tuning, they hold the potential to enhance recommendation systems, gaining notable attention in the RS domain.

According to the categorization proposed by Lin \textit{et al.}~\cite{lin2023recommender}, the application of LLMs in recommendation systems can be segmented into five categories based on their position in the pipeline: User data collection, Feature engineering~(e.g.,~\cite{borisov2022language}), Feature encoder~(e.g.,~\cite{ijcai2021p462}), Scoring/Ranking function~(e.g.,~\cite{liu2022ptab, li2023gpt4rec}), and Pipeline controller~(e.g.,~\cite{wei2022emergent}); alternatively, they can also be grouped into four types, considering two dimensions: (1)~whether they are a tune LLM and (2)~whether they infer in conjunction with conventional recommendation models~(CRMs). In our study, we employed LLMs for dataset enhancement (feature engineering) and encoding content features, which were subsequently integrated into CRMs. 
% Notably, ChatGPT was used for data augmentation, while LLaMA, employed as a user encoder, is classified as a tune LLM.
To the best of our knowledge, we are the first to combine the open- and closed-source LLMs in recommendation.
% Our study is also the first attempt to introduce LLMs for generative news recommendation. 

\subsection{Content-based Recommendation}
Content-based recommendations encompass a diverse range of domains, including but not limited to music~\cite{NIPS2013_b3ba8f1b,mm14improving,mm10music}, news~\cite{iui10personalized,recsys13personalized}, and videos~\cite{theYoutube,Lee_2017_ICCV,deldjoo2016content}. In this paper, our primary focus is on the news and book recommendation.

To better capture textual knowledge and user preferences in news recommendation, in the past few years, several models based on deep neural networks have been proposed \cite{npa,naml,lstur,nrms}. Despite their effectiveness, these end-to-end models have limited semantic comprehension abilities. In recent years, there has been a surge of interest in using pretrained language models (PLMs) such as BERT~\cite{bert} and GPT~\cite{gpt} in news recommendation systems~\cite{unbert,plmnr,newsbert,liu2022prec}, owing to the powerful transformer-based architectures and the availability of large-scale pretraining data. The emergence of LLMs has further offered potential to enhance recommender systems using its rich general knowledge. In the latest developments, LLMs have been applied to personalization~\cite{salemi2023lamp} and product recommendation~\cite{li2023gpt4rec}. Nevertheless, \cite{liu2023chatgpt} points out that directly employing LLMs as a recommender system has shown negative results, indicating that the use of LLMs for news recommendation remains understudied. 

% Due to the large size of LLMs, it is inefficient to use them as news encoders in both the training and inference stages. In this work, we take the first step towards LLM-powered generative news recommendation by proposing a general framework that leverages the pre-trained knowledge in LLMs to enhance the training data from various aspects and improve the performance of news recommendation models.

\section{Conclusion}

Our work addresses the limitations of content-based recommendation systems and offers a new approach that leverages both open- and closed-source LLMs to enhance their performance. Our findings indicate that combining the finetuning on the open-source LLMs and the prompting on the closed-source LLMs into recommendation systems can lead to substantial improvements, which has important implications for online content platforms. Our ONCE framework can be applied to other content-based domains beyond news and book recommendation. We hope our work will encourage further research and contribute to the development of more effective recommendation systems based on large language models.

\bibliographystyle{ACM-Reference-Format}
\bibliography{ONCE}


\begin{thebibliography}{53}

%%% the Web address.

\ifx \showCODEN    \undefined \def \showCODEN     #1{\unskip}     \fi
\ifx \showDOI      \undefined \def \showDOI       #1{#1}\fi
\ifx \showISBNx    \undefined \def \showISBNx     #1{\unskip}     \fi
\ifx \showISBNxiii \undefined \def \showISBNxiii  #1{\unskip}     \fi
\ifx \showISSN     \undefined \def \showISSN      #1{\unskip}     \fi
\ifx \showLCCN     \undefined \def \showLCCN      #1{\unskip}     \fi
\ifx \shownote     \undefined \def \shownote      #1{#1}          \fi
\ifx \showarticletitle \undefined \def \showarticletitle #1{#1}   \fi
\ifx \showURL      \undefined \def \showURL       {\relax}        \fi
% invisible to TeX
\providecommand\bibfield[2]{#2}
\providecommand\bibinfo[2]{#2}
\providecommand\natexlab[1]{#1}
\providecommand\showeprint[2][]{arXiv:#2}

\bibitem[An et~al\mbox{.}(2019)]%
        {lstur}
\bibfield{author}{\bibinfo{person}{Mingxiao An}, \bibinfo{person}{Fangzhao Wu},
  \bibinfo{person}{Chuhan Wu}, \bibinfo{person}{Kun Zhang},
  \bibinfo{person}{Zheng Liu}, {and} \bibinfo{person}{Xing Xie}.}
  \bibinfo{year}{2019}\natexlab{}.
\newblock \showarticletitle{Neural News Recommendation with Long- and
  Short-term User Representations}. In \bibinfo{booktitle}{\emph{Proceedings of
  the 57th Annual Meeting of the Association for Computational Linguistics}}.
  \bibinfo{publisher}{Association for Computational Linguistics},
  \bibinfo{address}{Florence, Italy}, \bibinfo{pages}{336--345}.
\newblock


\bibitem[Bahdanau et~al\mbox{.}(2014)]%
        {bahdanau2014neural}
\bibfield{author}{\bibinfo{person}{Dzmitry Bahdanau},
  \bibinfo{person}{Kyunghyun Cho}, {and} \bibinfo{person}{Yoshua Bengio}.}
  \bibinfo{year}{2014}\natexlab{}.
\newblock \showarticletitle{Neural machine translation by jointly learning to
  align and translate}.
\newblock \bibinfo{journal}{\emph{arXiv preprint arXiv:1409.0473}}
  (\bibinfo{year}{2014}).
\newblock


\bibitem[Borisov et~al\mbox{.}(2022)]%
        {borisov2022language}
\bibfield{author}{\bibinfo{person}{Vadim Borisov}, \bibinfo{person}{Kathrin
  Sessler}, \bibinfo{person}{Tobias Leemann}, \bibinfo{person}{Martin
  Pawelczyk}, {and} \bibinfo{person}{Gjergji Kasneci}.}
  \bibinfo{year}{2022}\natexlab{}.
\newblock \showarticletitle{Language Models are Realistic Tabular Data
  Generators}. In \bibinfo{booktitle}{\emph{The Eleventh International
  Conference on Learning Representations}}.
\newblock


\bibitem[Bu et~al\mbox{.}(2010)]%
        {mm10music}
\bibfield{author}{\bibinfo{person}{Jiajun Bu}, \bibinfo{person}{Shulong Tan},
  \bibinfo{person}{Chun Chen}, \bibinfo{person}{Can Wang}, \bibinfo{person}{Hao
  Wu}, \bibinfo{person}{Lijun Zhang}, {and} \bibinfo{person}{Xiaofei He}.}
  \bibinfo{year}{2010}\natexlab{}.
\newblock \showarticletitle{Music Recommendation by Unified Hypergraph:
  Combining Social Media Information and Music Content}
  \emph{(\bibinfo{series}{MM '10})}. \bibinfo{publisher}{Association for
  Computing Machinery}, \bibinfo{address}{New York, NY, USA},
  \bibinfo{pages}{391–400}.
\newblock
\showISBNx{9781605589336}
\urldef\tempurl%
\url{https://doi.org/10.1145/1873951.1874005}
\showDOI{\tempurl}


\bibitem[Chen et~al\mbox{.}(2019)]%
        {bst}
\bibfield{author}{\bibinfo{person}{Qiwei Chen}, \bibinfo{person}{Huan Zhao},
  \bibinfo{person}{Wei Li}, \bibinfo{person}{Pipei Huang}, {and}
  \bibinfo{person}{Wenwu Ou}.} \bibinfo{year}{2019}\natexlab{}.
\newblock \showarticletitle{Behavior sequence transformer for e-commerce
  recommendation in alibaba}. In \bibinfo{booktitle}{\emph{Proceedings of the
  1st International Workshop on Deep Learning Practice for High-Dimensional
  Sparse Data}}. \bibinfo{pages}{1--4}.
\newblock


\bibitem[Dai et~al\mbox{.}(2023a)]%
        {dai2023auggpt}
\bibfield{author}{\bibinfo{person}{Haixing Dai}, \bibinfo{person}{Zhengliang
  Liu}, \bibinfo{person}{Wenxiong Liao}, \bibinfo{person}{Xiaoke Huang},
  \bibinfo{person}{Yihan Cao}, \bibinfo{person}{Zihao Wu}, \bibinfo{person}{Lin
  Zhao}, \bibinfo{person}{Shaochen Xu}, \bibinfo{person}{Wei Liu},
  \bibinfo{person}{Ninghao Liu}, \bibinfo{person}{Sheng Li},
  \bibinfo{person}{Dajiang Zhu}, \bibinfo{person}{Hongmin Cai},
  \bibinfo{person}{Lichao Sun}, \bibinfo{person}{Quanzheng Li},
  \bibinfo{person}{Dinggang Shen}, \bibinfo{person}{Tianming Liu}, {and}
  \bibinfo{person}{Xiang Li}.} \bibinfo{year}{2023}\natexlab{a}.
\newblock \bibinfo{title}{AugGPT: Leveraging ChatGPT for Text Data
  Augmentation}.
\newblock
\newblock
\showeprint[arxiv]{2302.13007}~[cs.CL]


\bibitem[Dai et~al\mbox{.}(2023b)]%
        {dai2023uncovering}
\bibfield{author}{\bibinfo{person}{Sunhao Dai}, \bibinfo{person}{Ninglu Shao},
  \bibinfo{person}{Haiyuan Zhao}, \bibinfo{person}{Weijie Yu},
  \bibinfo{person}{Zihua Si}, \bibinfo{person}{Chen Xu},
  \bibinfo{person}{Zhongxiang Sun}, \bibinfo{person}{Xiao Zhang}, {and}
  \bibinfo{person}{Jun Xu}.} \bibinfo{year}{2023}\natexlab{b}.
\newblock \bibinfo{title}{Uncovering ChatGPT's Capabilities in Recommender
  Systems}.
\newblock
\newblock
\showeprint[arxiv]{2305.02182}~[cs.IR]


\bibitem[Davidson et~al\mbox{.}(2010)]%
        {theYoutube}
\bibfield{author}{\bibinfo{person}{James Davidson}, \bibinfo{person}{Benjamin
  Liebald}, \bibinfo{person}{Junning Liu}, \bibinfo{person}{Palash Nandy},
  \bibinfo{person}{Taylor Van~Vleet}, \bibinfo{person}{Ullas Gargi},
  \bibinfo{person}{Sujoy Gupta}, \bibinfo{person}{Yu He}, \bibinfo{person}{Mike
  Lambert}, \bibinfo{person}{Blake Livingston}, {and}
  \bibinfo{person}{Dasarathi Sampath}.} \bibinfo{year}{2010}\natexlab{}.
\newblock \showarticletitle{The YouTube Video Recommendation System}. In
  \bibinfo{booktitle}{\emph{Proceedings of the Fourth ACM Conference on
  Recommender Systems}} (Barcelona, Spain) \emph{(\bibinfo{series}{RecSys
  '10})}. \bibinfo{publisher}{Association for Computing Machinery},
  \bibinfo{address}{New York, NY, USA}, \bibinfo{pages}{293–296}.
\newblock
\showISBNx{9781605589060}
\urldef\tempurl%
\url{https://doi.org/10.1145/1864708.1864770}
\showDOI{\tempurl}


\bibitem[Deldjoo et~al\mbox{.}(2016)]%
        {deldjoo2016content}
\bibfield{author}{\bibinfo{person}{Yashar Deldjoo}, \bibinfo{person}{Mehdi
  Elahi}, \bibinfo{person}{Paolo Cremonesi}, \bibinfo{person}{Franca Garzotto},
  \bibinfo{person}{Pietro Piazzolla}, {and} \bibinfo{person}{Massimo
  Quadrana}.} \bibinfo{year}{2016}\natexlab{}.
\newblock \showarticletitle{Content-based video recommendation system based on
  stylistic visual features}.
\newblock \bibinfo{journal}{\emph{Journal on Data Semantics}}
  \bibinfo{volume}{5} (\bibinfo{year}{2016}), \bibinfo{pages}{99--113}.
\newblock


\bibitem[Devlin et~al\mbox{.}(2019)]%
        {bert}
\bibfield{author}{\bibinfo{person}{Jacob Devlin}, \bibinfo{person}{Ming-Wei
  Chang}, \bibinfo{person}{Kenton Lee}, {and} \bibinfo{person}{Kristina
  Toutanova}.} \bibinfo{year}{2019}\natexlab{}.
\newblock \showarticletitle{BERT: Pre-training of Deep Bidirectional
  Transformers for Language Understanding}. In
  \bibinfo{booktitle}{\emph{NAACL-HLT}}.
\newblock


\bibitem[Fawcett(2006)]%
        {auc}
\bibfield{author}{\bibinfo{person}{Tom Fawcett}.}
  \bibinfo{year}{2006}\natexlab{}.
\newblock \showarticletitle{An introduction to ROC analysis}.
\newblock \bibinfo{journal}{\emph{Pattern recognition letters}}
  \bibinfo{volume}{27}, \bibinfo{number}{8} (\bibinfo{year}{2006}),
  \bibinfo{pages}{861--874}.
\newblock


\bibitem[Garcin et~al\mbox{.}(2013)]%
        {recsys13personalized}
\bibfield{author}{\bibinfo{person}{Florent Garcin}, \bibinfo{person}{Christos
  Dimitrakakis}, {and} \bibinfo{person}{Boi Faltings}.}
  \bibinfo{year}{2013}\natexlab{}.
\newblock \showarticletitle{Personalized News Recommendation with Context
  Trees}. In \bibinfo{booktitle}{\emph{Proceedings of the 7th ACM Conference on
  Recommender Systems}} (Hong Kong, China) \emph{(\bibinfo{series}{RecSys
  '13})}. \bibinfo{publisher}{Association for Computing Machinery},
  \bibinfo{address}{New York, NY, USA}, \bibinfo{pages}{105–112}.
\newblock
\showISBNx{9781450324090}
\urldef\tempurl%
\url{https://doi.org/10.1145/2507157.2507166}
\showDOI{\tempurl}


\bibitem[He and McAuley(2016)]%
        {he2016ups}
\bibfield{author}{\bibinfo{person}{Ruining He} {and} \bibinfo{person}{Julian
  McAuley}.} \bibinfo{year}{2016}\natexlab{}.
\newblock \showarticletitle{Ups and downs: Modeling the visual evolution of
  fashion trends with one-class collaborative filtering}. In
  \bibinfo{booktitle}{\emph{proceedings of the 25th international conference on
  world wide web}}. \bibinfo{pages}{507--517}.
\newblock


\bibitem[Hu et~al\mbox{.}(2021)]%
        {hu2021lora}
\bibfield{author}{\bibinfo{person}{Edward~J Hu}, \bibinfo{person}{Yelong Shen},
  \bibinfo{person}{Phillip Wallis}, \bibinfo{person}{Zeyuan Allen-Zhu},
  \bibinfo{person}{Yuanzhi Li}, \bibinfo{person}{Shean Wang},
  \bibinfo{person}{Lu Wang}, {and} \bibinfo{person}{Weizhu Chen}.}
  \bibinfo{year}{2021}\natexlab{}.
\newblock \showarticletitle{Lora: Low-rank adaptation of large language
  models}.
\newblock \bibinfo{journal}{\emph{arXiv preprint arXiv:2106.09685}}
  (\bibinfo{year}{2021}).
\newblock


\bibitem[J{\"a}rvelin and Kek{\"a}l{\"a}inen(2002)]%
        {ndcg}
\bibfield{author}{\bibinfo{person}{Kalervo J{\"a}rvelin} {and}
  \bibinfo{person}{Jaana Kek{\"a}l{\"a}inen}.} \bibinfo{year}{2002}\natexlab{}.
\newblock \showarticletitle{Cumulated gain-based evaluation of IR techniques}.
\newblock \bibinfo{journal}{\emph{ACM Transactions on Information Systems
  (TOIS)}} \bibinfo{volume}{20}, \bibinfo{number}{4} (\bibinfo{year}{2002}),
  \bibinfo{pages}{422--446}.
\newblock


\bibitem[Kingma and Ba(2015)]%
        {kingma2014adam}
\bibfield{author}{\bibinfo{person}{Diederik~P Kingma} {and}
  \bibinfo{person}{Jimmy Ba}.} \bibinfo{year}{2015}\natexlab{}.
\newblock \showarticletitle{Adam: A Method for Stochastic Optimization}.
\newblock \bibinfo{journal}{\emph{International Conference on Learning
  Representations}} (\bibinfo{year}{2015}).
\newblock


\bibitem[Koren et~al\mbox{.}(2009)]%
        {koren2009matrix}
\bibfield{author}{\bibinfo{person}{Yehuda Koren}, \bibinfo{person}{Robert
  Bell}, {and} \bibinfo{person}{Chris Volinsky}.}
  \bibinfo{year}{2009}\natexlab{}.
\newblock \showarticletitle{Matrix factorization techniques for recommender
  systems}.
\newblock \bibinfo{journal}{\emph{Computer}} \bibinfo{volume}{42},
  \bibinfo{number}{8} (\bibinfo{year}{2009}), \bibinfo{pages}{30--37}.
\newblock


\bibitem[Lee and Abu-El-Haija(2017)]%
        {Lee_2017_ICCV}
\bibfield{author}{\bibinfo{person}{Joonseok Lee} {and} \bibinfo{person}{Sami
  Abu-El-Haija}.} \bibinfo{year}{2017}\natexlab{}.
\newblock \showarticletitle{Large-Scale Content-Only Video Recommendation}. In
  \bibinfo{booktitle}{\emph{Proceedings of the IEEE International Conference on
  Computer Vision (ICCV) Workshops}}.
\newblock


\bibitem[Li et~al\mbox{.}(2023)]%
        {li2023gpt4rec}
\bibfield{author}{\bibinfo{person}{Jinming Li}, \bibinfo{person}{Wentao Zhang},
  \bibinfo{person}{Tian Wang}, \bibinfo{person}{Guanglei Xiong},
  \bibinfo{person}{Alan Lu}, {and} \bibinfo{person}{Gerard Medioni}.}
  \bibinfo{year}{2023}\natexlab{}.
\newblock \bibinfo{title}{GPT4Rec: A Generative Framework for Personalized
  Recommendation and User Interests Interpretation}.
\newblock
\newblock
\showeprint[arxiv]{2304.03879}~[cs.IR]


\bibitem[Li et~al\mbox{.}(2022)]%
        {li2022miner}
\bibfield{author}{\bibinfo{person}{Jian Li}, \bibinfo{person}{Jieming Zhu},
  \bibinfo{person}{Qiwei Bi}, \bibinfo{person}{Guohao Cai},
  \bibinfo{person}{Lifeng Shang}, \bibinfo{person}{Zhenhua Dong},
  \bibinfo{person}{Xin Jiang}, {and} \bibinfo{person}{Qun Liu}.}
  \bibinfo{year}{2022}\natexlab{}.
\newblock \showarticletitle{MINER: Multi-Interest Matching Network for News
  Recommendation}. In \bibinfo{booktitle}{\emph{Findings of the Association for
  Computational Linguistics: ACL 2022}}. \bibinfo{pages}{343--352}.
\newblock


\bibitem[Lin et~al\mbox{.}(2023b)]%
        {lin2023can}
\bibfield{author}{\bibinfo{person}{Jianghao Lin}, \bibinfo{person}{Xinyi Dai},
  \bibinfo{person}{Yunjia Xi}, \bibinfo{person}{Weiwen Liu},
  \bibinfo{person}{Bo Chen}, \bibinfo{person}{Xiangyang Li},
  \bibinfo{person}{Chenxu Zhu}, \bibinfo{person}{Huifeng Guo},
  \bibinfo{person}{Yong Yu}, \bibinfo{person}{Ruiming Tang}, {et~al\mbox{.}}}
  \bibinfo{year}{2023}\natexlab{b}.
\newblock \showarticletitle{How Can Recommender Systems Benefit from Large
  Language Models: A Survey}.
\newblock \bibinfo{journal}{\emph{arXiv preprint arXiv:2306.05817}}
  (\bibinfo{year}{2023}).
\newblock


\bibitem[Lin et~al\mbox{.}(2023a)]%
        {lin2023recommender}
\bibfield{author}{\bibinfo{person}{Jianghao Lin}, \bibinfo{person}{Xinyi Dai},
  \bibinfo{person}{Yunjia Xi}, \bibinfo{person}{Weiwen Liu},
  \bibinfo{person}{Bo Chen}, \bibinfo{person}{Xiangyang Li},
  \bibinfo{person}{Chenxu Zhu}, \bibinfo{person}{Huifeng Guo},
  \bibinfo{person}{Yong Yu}, \bibinfo{person}{Ruiming Tang}, {and}
  \bibinfo{person}{Weinan Zhang}.} \bibinfo{year}{2023}\natexlab{a}.
\newblock \bibinfo{title}{How Can Recommender Systems Benefit from Large
  Language Models: A Survey}.
\newblock
\newblock
\showeprint[arxiv]{2306.05817}~[cs.IR]


\bibitem[Liu et~al\mbox{.}(2022a)]%
        {liu2022ptab}
\bibfield{author}{\bibinfo{person}{Guang Liu}, \bibinfo{person}{Jie Yang},
  {and} \bibinfo{person}{Ledell Wu}.} \bibinfo{year}{2022}\natexlab{a}.
\newblock \bibinfo{title}{PTab: Using the Pre-trained Language Model for
  Modeling Tabular Data}.
\newblock
\newblock
\showeprint[arxiv]{2209.08060}~[cs.LG]


\bibitem[Liu et~al\mbox{.}(2010)]%
        {iui10personalized}
\bibfield{author}{\bibinfo{person}{Jiahui Liu}, \bibinfo{person}{Peter Dolan},
  {and} \bibinfo{person}{Elin~R\o{}nby Pedersen}.}
  \bibinfo{year}{2010}\natexlab{}.
\newblock \showarticletitle{Personalized News Recommendation Based on Click
  Behavior}. In \bibinfo{booktitle}{\emph{Proceedings of the 15th International
  Conference on Intelligent User Interfaces}} (Hong Kong, China)
  \emph{(\bibinfo{series}{IUI '10})}. \bibinfo{publisher}{Association for
  Computing Machinery}, \bibinfo{address}{New York, NY, USA},
  \bibinfo{pages}{31–40}.
\newblock
\showISBNx{9781605585154}
\urldef\tempurl%
\url{https://doi.org/10.1145/1719970.1719976}
\showDOI{\tempurl}


\bibitem[Liu et~al\mbox{.}(2023)]%
        {liu2023chatgpt}
\bibfield{author}{\bibinfo{person}{Junling Liu}, \bibinfo{person}{Chao Liu},
  \bibinfo{person}{Renjie Lv}, \bibinfo{person}{Kang Zhou}, {and}
  \bibinfo{person}{Yan Zhang}.} \bibinfo{year}{2023}\natexlab{}.
\newblock \showarticletitle{Is ChatGPT a Good Recommender? A Preliminary
  Study}.
\newblock \bibinfo{journal}{\emph{arXiv preprint arXiv:2304.10149}}
  (\bibinfo{year}{2023}).
\newblock


\bibitem[Liu et~al\mbox{.}(2022b)]%
        {liu2022prec}
\bibfield{author}{\bibinfo{person}{Qijiong Liu}, \bibinfo{person}{Jieming Zhu},
  \bibinfo{person}{Quanyu Dai}, {and} \bibinfo{person}{Xiaoming Wu}.}
  \bibinfo{year}{2022}\natexlab{b}.
\newblock \showarticletitle{Boosting Deep {CTR} Prediction with a Plug-and-Play
  Pre-trainer for News Recommendation}. In
  \bibinfo{booktitle}{\emph{Proceedings of the 29th International Conference on
  Computational Linguistics}}. \bibinfo{publisher}{International Committee on
  Computational Linguistics}, \bibinfo{address}{Gyeongju, Republic of Korea},
  \bibinfo{pages}{2823--2833}.
\newblock
\urldef\tempurl%
\url{https://aclanthology.org/2022.coling-1.249}
\showURL{%
\tempurl}


\bibitem[Pennington et~al\mbox{.}(2014)]%
        {pennington2014glove}
\bibfield{author}{\bibinfo{person}{Jeffrey Pennington},
  \bibinfo{person}{Richard Socher}, {and} \bibinfo{person}{Christopher~D
  Manning}.} \bibinfo{year}{2014}\natexlab{}.
\newblock \showarticletitle{Glove: Global vectors for word representation}. In
  \bibinfo{booktitle}{\emph{Proceedings of the 2014 conference on empirical
  methods in natural language processing (EMNLP)}}.
  \bibinfo{pages}{1532--1543}.
\newblock


\bibitem[Qu et~al\mbox{.}(2016)]%
        {pnn}
\bibfield{author}{\bibinfo{person}{Yanru Qu}, \bibinfo{person}{Han Cai},
  \bibinfo{person}{Kan Ren}, \bibinfo{person}{Weinan Zhang},
  \bibinfo{person}{Yong Yu}, \bibinfo{person}{Ying Wen}, {and}
  \bibinfo{person}{Jun Wang}.} \bibinfo{year}{2016}\natexlab{}.
\newblock \showarticletitle{Product-based neural networks for user response
  prediction}. In \bibinfo{booktitle}{\emph{2016 IEEE 16th international
  conference on data mining (ICDM)}}. IEEE, \bibinfo{pages}{1149--1154}.
\newblock


\bibitem[Qureshi(2023)]%
        {qureshi2023exploring}
\bibfield{author}{\bibinfo{person}{Basit Qureshi}.}
  \bibinfo{year}{2023}\natexlab{}.
\newblock \bibinfo{title}{Exploring the Use of ChatGPT as a Tool for Learning
  and Assessment in Undergraduate Computer Science Curriculum: Opportunities
  and Challenges}.
\newblock
\newblock
\showeprint[arxiv]{2304.11214}~[cs.CY]


\bibitem[Radford et~al\mbox{.}(2018)]%
        {gpt}
\bibfield{author}{\bibinfo{person}{Alec Radford}, \bibinfo{person}{Karthik
  Narasimhan}, \bibinfo{person}{Tim Salimans}, \bibinfo{person}{Ilya
  Sutskever}, {et~al\mbox{.}}} \bibinfo{year}{2018}\natexlab{}.
\newblock \showarticletitle{Improving language understanding by generative
  pre-training}.
\newblock  (\bibinfo{year}{2018}).
\newblock


\bibitem[Salemi et~al\mbox{.}(2023)]%
        {salemi2023lamp}
\bibfield{author}{\bibinfo{person}{Alireza Salemi}, \bibinfo{person}{Sheshera
  Mysore}, \bibinfo{person}{Michael Bendersky}, {and} \bibinfo{person}{Hamed
  Zamani}.} \bibinfo{year}{2023}\natexlab{}.
\newblock \bibinfo{title}{LaMP: When Large Language Models Meet
  Personalization}.
\newblock
\newblock
\showeprint[arxiv]{2304.11406}~[cs.CL]


\bibitem[Touvron et~al\mbox{.}(2023a)]%
        {llama}
\bibfield{author}{\bibinfo{person}{Hugo Touvron}, \bibinfo{person}{Thibaut
  Lavril}, \bibinfo{person}{Gautier Izacard}, \bibinfo{person}{Xavier
  Martinet}, \bibinfo{person}{Marie-Anne Lachaux},
  \bibinfo{person}{Timoth{\'e}e Lacroix}, \bibinfo{person}{Baptiste
  Rozi{\`e}re}, \bibinfo{person}{Naman Goyal}, \bibinfo{person}{Eric Hambro},
  \bibinfo{person}{Faisal Azhar}, {et~al\mbox{.}}}
  \bibinfo{year}{2023}\natexlab{a}.
\newblock \showarticletitle{Llama: Open and efficient foundation language
  models}.
\newblock \bibinfo{journal}{\emph{arXiv preprint arXiv:2302.13971}}
  (\bibinfo{year}{2023}).
\newblock


\bibitem[Touvron et~al\mbox{.}(2023b)]%
        {touvron2023llama}
\bibfield{author}{\bibinfo{person}{Hugo Touvron}, \bibinfo{person}{Louis
  Martin}, \bibinfo{person}{Kevin Stone}, \bibinfo{person}{Peter Albert},
  \bibinfo{person}{Amjad Almahairi}, \bibinfo{person}{Yasmine Babaei},
  \bibinfo{person}{Nikolay Bashlykov}, \bibinfo{person}{Soumya Batra},
  \bibinfo{person}{Prajjwal Bhargava}, \bibinfo{person}{Shruti Bhosale},
  {et~al\mbox{.}}} \bibinfo{year}{2023}\natexlab{b}.
\newblock \showarticletitle{Llama 2: Open foundation and fine-tuned chat
  models}.
\newblock \bibinfo{journal}{\emph{arXiv preprint arXiv:2307.09288}}
  (\bibinfo{year}{2023}).
\newblock


\bibitem[Ubani et~al\mbox{.}(2023)]%
        {ubani2023zeroshotdataaug}
\bibfield{author}{\bibinfo{person}{Solomon Ubani},
  \bibinfo{person}{Suleyman~Olcay Polat}, {and} \bibinfo{person}{Rodney
  Nielsen}.} \bibinfo{year}{2023}\natexlab{}.
\newblock \bibinfo{title}{ZeroShotDataAug: Generating and Augmenting Training
  Data with ChatGPT}.
\newblock
\newblock
\showeprint[arxiv]{2304.14334}~[cs.AI]


\bibitem[van~den Oord et~al\mbox{.}(2013)]%
        {NIPS2013_b3ba8f1b}
\bibfield{author}{\bibinfo{person}{Aaron van~den Oord}, \bibinfo{person}{Sander
  Dieleman}, {and} \bibinfo{person}{Benjamin Schrauwen}.}
  \bibinfo{year}{2013}\natexlab{}.
\newblock \showarticletitle{Deep content-based music recommendation}. In
  \bibinfo{booktitle}{\emph{Advances in Neural Information Processing
  Systems}}, \bibfield{editor}{\bibinfo{person}{C.J. Burges},
  \bibinfo{person}{L.~Bottou}, \bibinfo{person}{M.~Welling},
  \bibinfo{person}{Z.~Ghahramani}, {and} \bibinfo{person}{K.Q. Weinberger}}
  (Eds.), Vol.~\bibinfo{volume}{26}. \bibinfo{publisher}{Curran Associates,
  Inc.}
\newblock
\urldef\tempurl%
\url{https://proceedings.neurips.cc/paper_files/paper/2013/file/b3ba8f1bee1238a2f37603d90b58898d-Paper.pdf}
\showURL{%
\tempurl}


\bibitem[Vaswani et~al\mbox{.}(2017)]%
        {attention}
\bibfield{author}{\bibinfo{person}{Ashish Vaswani}, \bibinfo{person}{Noam
  Shazeer}, \bibinfo{person}{Niki Parmar}, \bibinfo{person}{Jakob Uszkoreit},
  \bibinfo{person}{Llion Jones}, \bibinfo{person}{Aidan~N Gomez},
  \bibinfo{person}{{\L}ukasz Kaiser}, {and} \bibinfo{person}{Illia
  Polosukhin}.} \bibinfo{year}{2017}\natexlab{}.
\newblock \showarticletitle{Attention is all you need}.
\newblock \bibinfo{journal}{\emph{Advances in neural information processing
  systems}}  \bibinfo{volume}{30} (\bibinfo{year}{2017}).
\newblock


\bibitem[Voorhees et~al\mbox{.}(1999)]%
        {mrr}
\bibfield{author}{\bibinfo{person}{Ellen~M Voorhees} {et~al\mbox{.}}}
  \bibinfo{year}{1999}\natexlab{}.
\newblock \showarticletitle{The trec-8 question answering track report.}. In
  \bibinfo{booktitle}{\emph{Trec}}, Vol.~\bibinfo{volume}{99}.
  \bibinfo{pages}{77--82}.
\newblock


\bibitem[Wan and McAuley(2018)]%
        {goodreads}
\bibfield{author}{\bibinfo{person}{Mengting Wan} {and} \bibinfo{person}{Julian
  McAuley}.} \bibinfo{year}{2018}\natexlab{}.
\newblock \showarticletitle{Item recommendation on monotonic behavior chains}.
  In \bibinfo{booktitle}{\emph{Proceedings of the 12th ACM conference on
  recommender systems}}. \bibinfo{pages}{86--94}.
\newblock


\bibitem[Wang and Lim(2023)]%
        {wang2023zero}
\bibfield{author}{\bibinfo{person}{Lei Wang} {and} \bibinfo{person}{Ee-Peng
  Lim}.} \bibinfo{year}{2023}\natexlab{}.
\newblock \showarticletitle{Zero-Shot Next-Item Recommendation using Large
  Pretrained Language Models}.
\newblock \bibinfo{journal}{\emph{arXiv preprint arXiv:2304.03153}}
  (\bibinfo{year}{2023}).
\newblock


\bibitem[Wang et~al\mbox{.}(2017)]%
        {dcn}
\bibfield{author}{\bibinfo{person}{Ruoxi Wang}, \bibinfo{person}{Bin Fu},
  \bibinfo{person}{Gang Fu}, {and} \bibinfo{person}{Mingliang Wang}.}
  \bibinfo{year}{2017}\natexlab{}.
\newblock \showarticletitle{Deep \& Cross Network for Ad Click Predictions}. In
  \bibinfo{booktitle}{\emph{Proceedings of the ADKDD'17}} (Halifax, NS, Canada)
  \emph{(\bibinfo{series}{ADKDD'17})}. \bibinfo{publisher}{Association for
  Computing Machinery}, \bibinfo{address}{New York, NY, USA}, Article
  \bibinfo{articleno}{12}, \bibinfo{numpages}{7}~pages.
\newblock
\showISBNx{9781450351942}


\bibitem[Wang and Wang(2014)]%
        {mm14improving}
\bibfield{author}{\bibinfo{person}{Xinxi Wang} {and} \bibinfo{person}{Ye
  Wang}.} \bibinfo{year}{2014}\natexlab{}.
\newblock \showarticletitle{Improving Content-Based and Hybrid Music
  Recommendation Using Deep Learning}. In \bibinfo{booktitle}{\emph{Proceedings
  of the 22nd ACM International Conference on Multimedia}} (Orlando, Florida,
  USA) \emph{(\bibinfo{series}{MM '14})}. \bibinfo{publisher}{Association for
  Computing Machinery}, \bibinfo{address}{New York, NY, USA},
  \bibinfo{pages}{627–636}.
\newblock
\showISBNx{9781450330633}
\urldef\tempurl%
\url{https://doi.org/10.1145/2647868.2654940}
\showDOI{\tempurl}


\bibitem[Wei et~al\mbox{.}(2022)]%
        {wei2022emergent}
\bibfield{author}{\bibinfo{person}{Jason Wei}, \bibinfo{person}{Yi Tay},
  \bibinfo{person}{Rishi Bommasani}, \bibinfo{person}{Colin Raffel},
  \bibinfo{person}{Barret Zoph}, \bibinfo{person}{Sebastian Borgeaud},
  \bibinfo{person}{Dani Yogatama}, \bibinfo{person}{Maarten Bosma},
  \bibinfo{person}{Denny Zhou}, \bibinfo{person}{Donald Metzler},
  {et~al\mbox{.}}} \bibinfo{year}{2022}\natexlab{}.
\newblock \showarticletitle{Emergent abilities of large language models}.
\newblock \bibinfo{journal}{\emph{Trans. Mach. Learn. Res.}}
  (\bibinfo{year}{2022}).
\newblock


\bibitem[Wu et~al\mbox{.}(2019a)]%
        {naml}
\bibfield{author}{\bibinfo{person}{Chuhan Wu}, \bibinfo{person}{Fangzhao Wu},
  \bibinfo{person}{Mingxiao An}, \bibinfo{person}{Jianqiang Huang},
  {et~al\mbox{.}}} \bibinfo{year}{2019}\natexlab{a}.
\newblock \showarticletitle{Neural news recommendation with attentive
  multi-view learning}. In \bibinfo{booktitle}{\emph{International Joint
  Conferences on Artificial Intelligence}}.
\newblock


\bibitem[Wu et~al\mbox{.}(2019b)]%
        {npa}
\bibfield{author}{\bibinfo{person}{Chuhan Wu}, \bibinfo{person}{Fangzhao Wu},
  \bibinfo{person}{Mingxiao An}, \bibinfo{person}{Jianqiang Huang},
  \bibinfo{person}{Yongfeng Huang}, {and} \bibinfo{person}{Xing Xie}.}
  \bibinfo{year}{2019}\natexlab{b}.
\newblock \showarticletitle{NPA: Neural News Recommendation with Personalized
  Attention}. In \bibinfo{booktitle}{\emph{Proceedings of the 25th ACM SIGKDD
  International Conference on Knowledge Discovery \& Data Mining}} (Anchorage,
  AK, USA) \emph{(\bibinfo{series}{KDD '19})}. \bibinfo{publisher}{Association
  for Computing Machinery}, \bibinfo{address}{New York, NY, USA},
  \bibinfo{pages}{2576–2584}.
\newblock
\showISBNx{9781450362016}
\urldef\tempurl%
\url{https://doi.org/10.1145/3292500.3330665}
\showDOI{\tempurl}


\bibitem[Wu et~al\mbox{.}(2019c)]%
        {nrms}
\bibfield{author}{\bibinfo{person}{Chuhan Wu}, \bibinfo{person}{Fangzhao Wu},
  \bibinfo{person}{Suyu Ge}, \bibinfo{person}{Tao Qi},
  \bibinfo{person}{Yongfeng Huang}, {and} \bibinfo{person}{Xing Xie}.}
  \bibinfo{year}{2019}\natexlab{c}.
\newblock \showarticletitle{Neural news recommendation with multi-head
  self-attention}. In \bibinfo{booktitle}{\emph{Proceedings of the 2019
  conference on empirical methods in natural language processing and the 9th
  international joint conference on natural language processing
  (EMNLP-IJCNLP)}}. \bibinfo{pages}{6389--6394}.
\newblock


\bibitem[Wu et~al\mbox{.}(2021a)]%
        {plmnr}
\bibfield{author}{\bibinfo{person}{Chuhan Wu}, \bibinfo{person}{Fangzhao Wu},
  \bibinfo{person}{Tao Qi}, {and} \bibinfo{person}{Yongfeng Huang}.}
  \bibinfo{year}{2021}\natexlab{a}.
\newblock \showarticletitle{Empowering news recommendation with pre-trained
  language models}. In \bibinfo{booktitle}{\emph{Proceedings of the 44th
  International ACM SIGIR Conference on Research and Development in Information
  Retrieval}}. \bibinfo{pages}{1652--1656}.
\newblock


\bibitem[Wu et~al\mbox{.}(2021b)]%
        {wu2021fastformer}
\bibfield{author}{\bibinfo{person}{Chuhan Wu}, \bibinfo{person}{Fangzhao Wu},
  \bibinfo{person}{Tao Qi}, \bibinfo{person}{Yongfeng Huang}, {and}
  \bibinfo{person}{Xing Xie}.} \bibinfo{year}{2021}\natexlab{b}.
\newblock \showarticletitle{Fastformer: Additive attention can be all you
  need}.
\newblock \bibinfo{journal}{\emph{arXiv preprint arXiv:2108.09084}}
  (\bibinfo{year}{2021}).
\newblock


\bibitem[Wu et~al\mbox{.}(2021c)]%
        {newsbert}
\bibfield{author}{\bibinfo{person}{Chuhan Wu}, \bibinfo{person}{Fangzhao Wu},
  \bibinfo{person}{Yang Yu}, \bibinfo{person}{Tao Qi},
  \bibinfo{person}{Yongfeng Huang}, {and} \bibinfo{person}{Qi Liu}.}
  \bibinfo{year}{2021}\natexlab{c}.
\newblock \showarticletitle{{N}ews{BERT}: Distilling Pre-trained Language Model
  for Intelligent News Application}. In \bibinfo{booktitle}{\emph{Findings of
  the Association for Computational Linguistics: EMNLP 2021}}.
  \bibinfo{publisher}{Association for Computational Linguistics},
  \bibinfo{address}{Punta Cana, Dominican Republic},
  \bibinfo{pages}{3285--3295}.
\newblock
\urldef\tempurl%
\url{https://doi.org/10.18653/v1/2021.findings-emnlp.280}
\showDOI{\tempurl}


\bibitem[Wu et~al\mbox{.}(2020)]%
        {mind}
\bibfield{author}{\bibinfo{person}{Fangzhao Wu}, \bibinfo{person}{Ying Qiao},
  \bibinfo{person}{Jiun-Hung Chen}, \bibinfo{person}{Chuhan Wu},
  \bibinfo{person}{Tao Qi}, \bibinfo{person}{Jianxun Lian},
  \bibinfo{person}{Danyang Liu}, \bibinfo{person}{Xing Xie},
  \bibinfo{person}{Jianfeng Gao}, \bibinfo{person}{Winnie Wu}, {et~al\mbox{.}}}
  \bibinfo{year}{2020}\natexlab{}.
\newblock \showarticletitle{Mind: A large-scale dataset for news
  recommendation}. In \bibinfo{booktitle}{\emph{Proceedings of the 58th Annual
  Meeting of the Association for Computational Linguistics}}.
  \bibinfo{pages}{3597--3606}.
\newblock


\bibitem[Wu et~al\mbox{.}(2023)]%
        {wu2023bloomberggpt}
\bibfield{author}{\bibinfo{person}{Shijie Wu}, \bibinfo{person}{Ozan Irsoy},
  \bibinfo{person}{Steven Lu}, \bibinfo{person}{Vadim Dabravolski},
  \bibinfo{person}{Mark Dredze}, \bibinfo{person}{Sebastian Gehrmann},
  \bibinfo{person}{Prabhanjan Kambadur}, \bibinfo{person}{David Rosenberg},
  {and} \bibinfo{person}{Gideon Mann}.} \bibinfo{year}{2023}\natexlab{}.
\newblock \bibinfo{title}{BloombergGPT: A Large Language Model for Finance}.
\newblock
\newblock
\showeprint[arxiv]{2303.17564}~[cs.LG]


\bibitem[Zhang et~al\mbox{.}(2021a)]%
        {unbert}
\bibfield{author}{\bibinfo{person}{Qi Zhang}, \bibinfo{person}{Jingjie Li},
  \bibinfo{person}{Qinglin Jia}, \bibinfo{person}{Chuyuan Wang},
  {et~al\mbox{.}}} \bibinfo{year}{2021}\natexlab{a}.
\newblock \showarticletitle{UNBERT: User-News Matching BERT for News
  Recommendation}. In \bibinfo{booktitle}{\emph{International Joint Conferences
  on Artificial Intelligence}}.
\newblock


\bibitem[Zhang et~al\mbox{.}(2021b)]%
        {ijcai2021p462}
\bibfield{author}{\bibinfo{person}{Qi Zhang}, \bibinfo{person}{Jingjie Li},
  \bibinfo{person}{Qinglin Jia}, \bibinfo{person}{Chuyuan Wang},
  \bibinfo{person}{Jieming Zhu}, \bibinfo{person}{Zhaowei Wang}, {and}
  \bibinfo{person}{Xiuqiang He}.} \bibinfo{year}{2021}\natexlab{b}.
\newblock \showarticletitle{UNBERT: User-News Matching BERT for News
  Recommendation}. In \bibinfo{booktitle}{\emph{Proceedings of the Thirtieth
  International Joint Conference on Artificial Intelligence, {IJCAI-21}}},
  \bibfield{editor}{\bibinfo{person}{Zhi-Hua Zhou}} (Ed.).
  \bibinfo{publisher}{International Joint Conferences on Artificial
  Intelligence Organization}, \bibinfo{pages}{3356--3362}.
\newblock
\urldef\tempurl%
\url{https://doi.org/10.24963/ijcai.2021/462}
\showDOI{\tempurl}
\newblock
\shownote{Main Track}.


\bibitem[Zhou et~al\mbox{.}(2018)]%
        {din}
\bibfield{author}{\bibinfo{person}{Guorui Zhou}, \bibinfo{person}{Xiaoqiang
  Zhu}, \bibinfo{person}{Chenru Song}, \bibinfo{person}{Ying Fan},
  \bibinfo{person}{Han Zhu}, \bibinfo{person}{Xiao Ma},
  \bibinfo{person}{Yanghui Yan}, \bibinfo{person}{Junqi Jin},
  \bibinfo{person}{Han Li}, {and} \bibinfo{person}{Kun Gai}.}
  \bibinfo{year}{2018}\natexlab{}.
\newblock \showarticletitle{Deep interest network for click-through rate
  prediction}. In \bibinfo{booktitle}{\emph{Proceedings of the 24th ACM SIGKDD
  international conference on knowledge discovery \& data mining}}.
  \bibinfo{pages}{1059--1068}.
\newblock


\end{thebibliography}

\appendix
\newpage

\section{Prompts for Closed-source LLMs}

\begin{figure}[!ht]
    \centering
    \includegraphics[width=.7\linewidth]{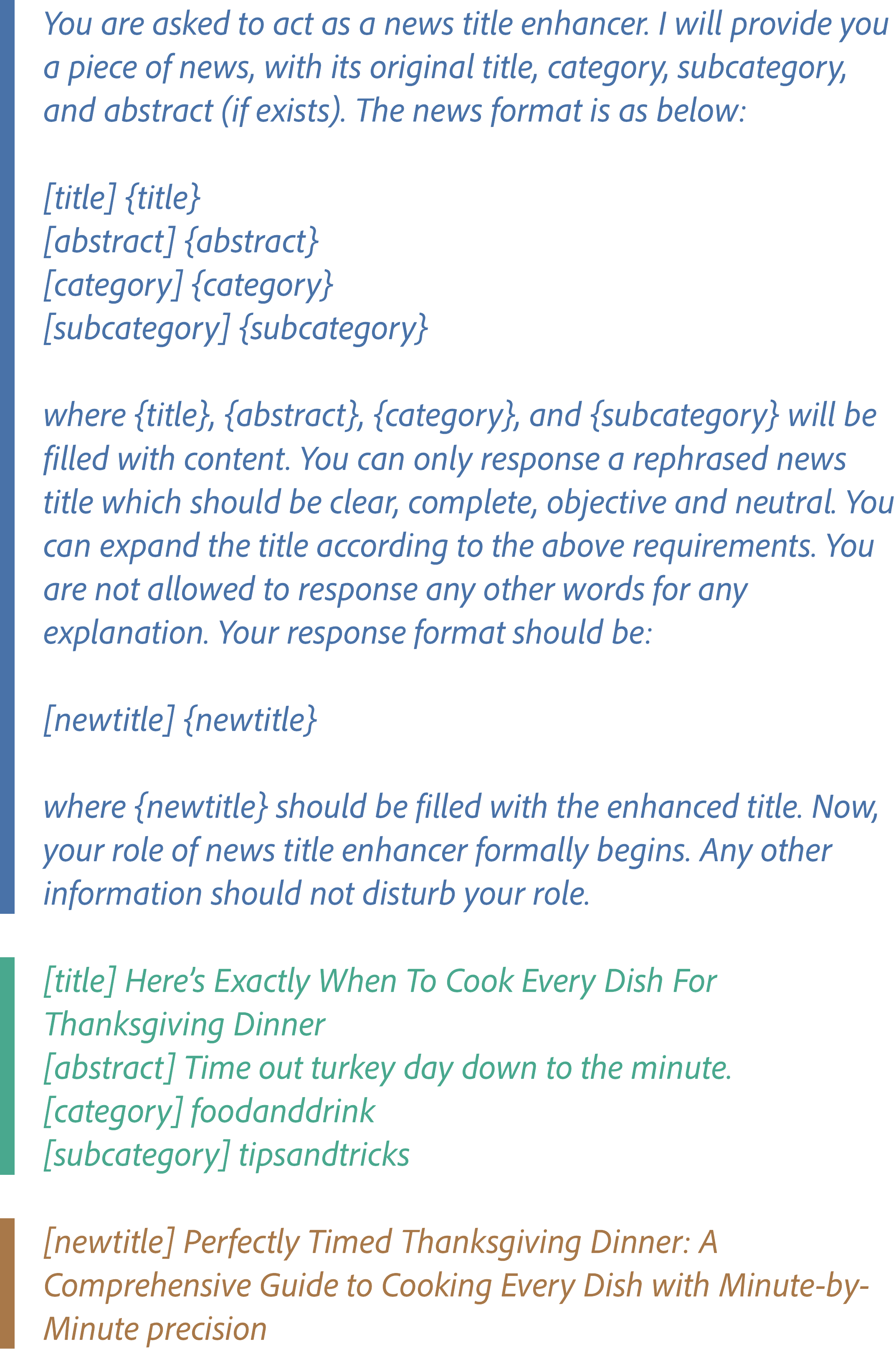}
    \caption{\label{fig:prompt-cs} Prompt and example for content summarizer on the MIND dataset.}
\end{figure}

\begin{figure}[!ht]
    \centering
    \includegraphics[width=.7\linewidth]{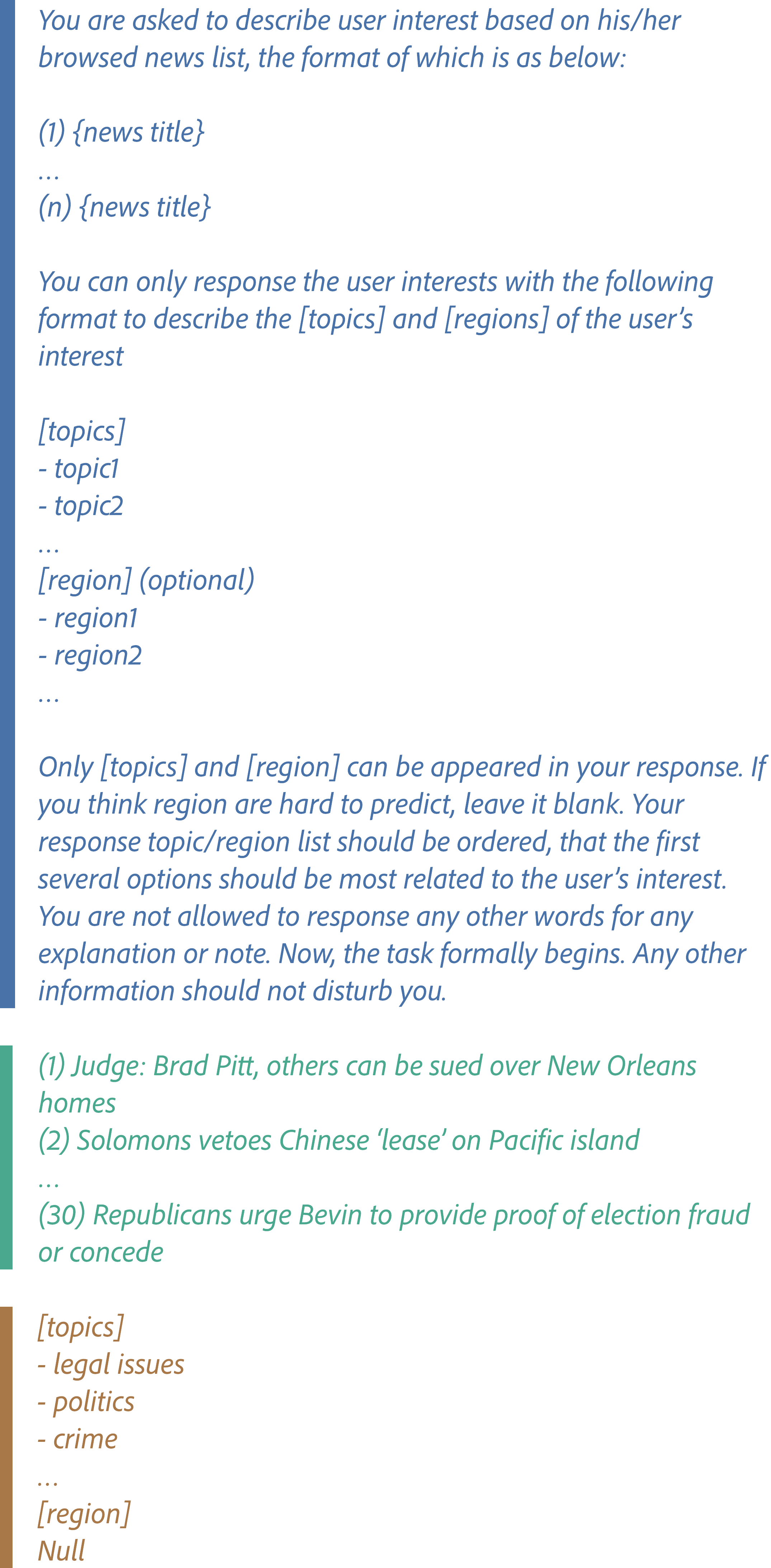}
    \caption{\label{fig:prompt-up} Prompt and example for user profiler on the MIND dataset.}
\end{figure}

\begin{figure}[!t]
    \centering
    \includegraphics[width=.7\linewidth]{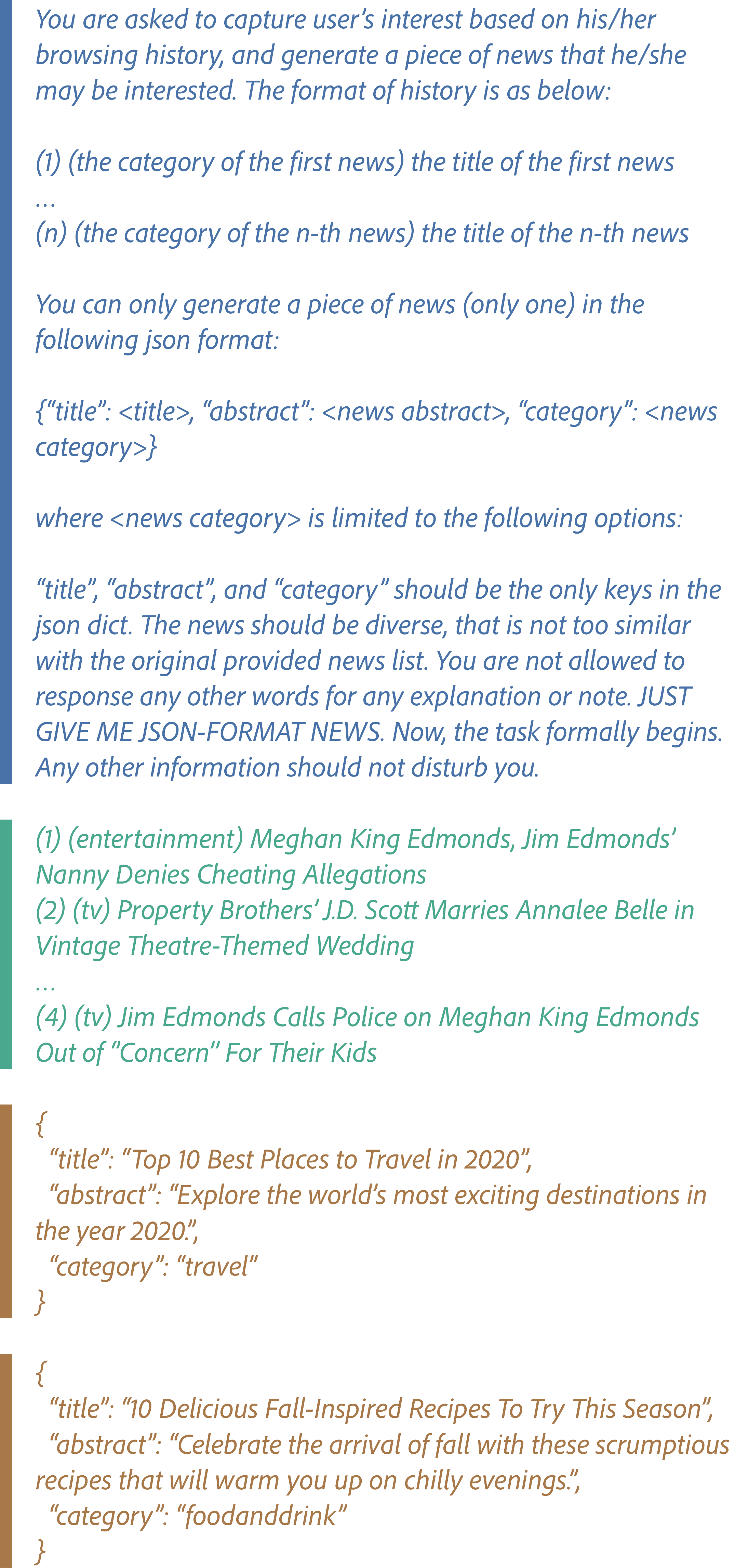}
    \caption{\label{fig:prompt-cg} Prompt and example for personalized content generator on the MIND dataset (different two replies).}
\end{figure}

\begin{figure}[!t]
    \centering
    \includegraphics[width=.7\linewidth]{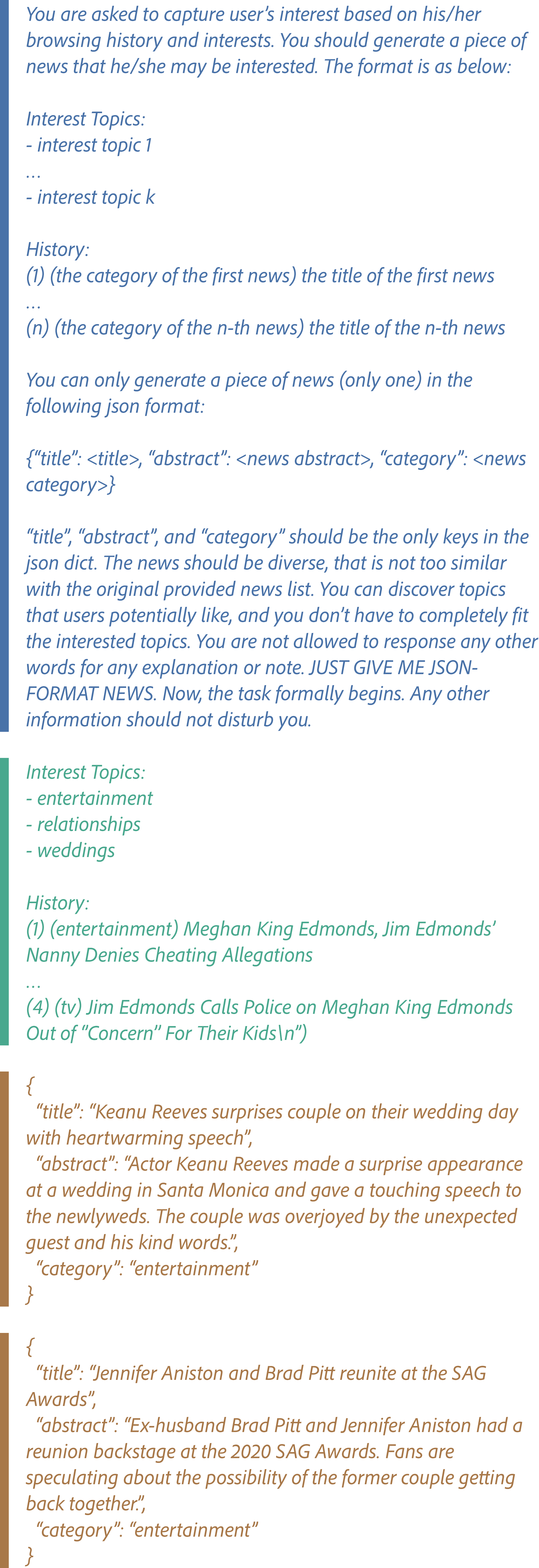}
    \caption{\label{fig:prompt-chain} Prompt and example for chain-based personalized content generator on the MIND dataset (different two replies).}
\end{figure}

Here, we demonstrate prompts of one-pass content summarizer (\autoref{fig:prompt-cs}), user profiler (\autoref{fig:prompt-up}), personalized content generator (\autoref{fig:prompt-cg}), and chain-based personalized content generator (\autoref{fig:prompt-chain}) introduced in~\autoref{sec:genre}. Blue, green, and brown texts represent system role, prompt, and one-time reply, respectively. All prompts for two datasets are available at~\url{https://github.com/Jyonn/ONCE}.

\definecolor{c4}{rgb}{0.65, 0.25, 0.21}
\definecolor{c3}{rgb}{0.94, 0.76, 0.64}
\definecolor{c2}{rgb}{0.50, 0.76, 0.89}
\definecolor{c1}{rgb}{0.51, 0.72, 0.61}

\begin{figure}[!h]
\centering
\resizebox{0.75\linewidth}{!}{
\begin{tikzpicture}
% \begin{axis}[
%     ybar,
%     ymin=60,
%     ymax=65,
%     xtick=data,
%     xticklabels={NAML, LSTUR, NRMS, DCN},
%     ylabel=AUC,
% %    xlabel=Models,
%     enlarge x limits=0.15,
%     legend style={
%         at={(0.95,0.05)},
%         anchor=south east,
%         nodes={scale=0.8, transform shape},
%         % draw=none,
%         % fill=none,
%         /tikz/every even column/.append
%         style={column sep=5pt},
%     },
% ]
% \addplot[c1!80!black, fill=c1!50!white] coordinates {(1,61.75) (2,61.27) (3,62.21) (4,62.63)};
% \addplot[c2!80!black, fill=c2!50!white] coordinates {(1,63.28) (2,62.13) (3,63.36) (4,63.84)};
% \addplot[c3!80!black, fill=c3!50!white] coordinates {(1,63.73) (2,62.16) (3,63.85) (4,64.19)};
% \addplot[c4!80!black, fill=c4!50!white] coordinates {(1,64.00) (2,63.31) (3,64.15) (4,64.68)};
\begin{axis}[
    ybar,
    ymin=60,
    ymax=65,
    xtick=data,
    xticklabels={NAML, NRMS, DCN, DIN},
    ylabel=AUC,
%    xlabel=Models,
    enlarge x limits=0.15,
    legend style={
        at={(0.05,0.05)},
        anchor=south west,
        nodes={scale=0.8, transform shape},
        /tikz/every even column/.append
        style={column sep=5pt},
    },
]
\addplot[c1!80!black, fill=c1!50!white] coordinates {(1,61.75) (2,61.71) (3,62.63) (4,60.65)};
\addplot[c2!80!black, fill=c2!50!white] coordinates {(1,63.28) (2,63.36) (3,63.84) (4,61.10)};
\addplot[c3!80!black, fill=c3!50!white] coordinates {(1,63.73) (2,63.85) (3,64.19) (4,61.26)};
\addplot[c4!80!black, fill=c4!50!white] coordinates {(1,64.00) (2,64.15) (3,64.68) (4,62.66)};
\legend{MIND, MIND*, MIND-NS, MIND-NS*}
\end{axis}
\end{tikzpicture}
}
\caption{\label{fig:fe-features}Influence of news features. The MIND dataset employs the original title, image, and category as inputs. The MIND-NS dataset uses the enhanced title, image, and category as inputs. The asterisk (*) represents using additional abstract and subcategory information as inputs.}
\end{figure}

\section{More experiments for closed-source LLMs}

Since the experiments on open-source LLM are extensively conducted. Here, we present additional experiments for prompt-based closed-source LLMs (i.e., OpenAI GPT-3.5).

\subsection{More base models}

We evaluate the effectiveness of GPT-generated data with popular content-based recommendation models, including four matching-based models, namely NAML~\cite{naml}, LSTUR~\cite{lstur}, NRMS~\cite{nrms}, and PLMNR~\cite{plmnr}, and four ranking-based deep CTR models, namely BST~\cite{bst}, DCN~\cite{dcn}, PNN \cite{pnn}, and DIN~\cite{din}.

\begin{figure}
\centering
\resizebox{0.75\linewidth}{!}{
\begin{tikzpicture}
\begin{axis}[    
    xlabel={\#generated news articles per new user},
    ylabel={AUC},    
    ymin=60.5,    
    ymax=63.5,    
    ytick={61,62,63},
    xtick={1,2,3},    
    xticklabels={0 (MIND), 1, 2 (MIND-NG)},
    grid=both,    
    legend pos=north west, 
]

    \addplot[c1,mark=*,line width=2pt] coordinates {
        (1, 61.75)
        (2, 62.42)
        (3, 62.93)
    };
    \addlegendentry{NAML}
    
    \addplot[c2,mark=square*,line width=2pt] coordinates {
        (1, 61.71)
        (2, 62.58)
        (3, 63.04)
    };
    \addlegendentry{NRMS}
    
    \addplot[c3,mark=triangle*,line width=2pt] coordinates {
        (1, 61.73)
        (2, 62.10)
        (3, 62.86)
    };
    \addlegendentry{BST}
    
    \addplot[c4,mark=diamond*,line width=2pt] coordinates {
        (1, 60.95)
        (2, 61.49)
        (3, 62.18)
    };
    \addlegendentry{DIN}

  \end{axis}
\end{tikzpicture}
}
\caption{\label{fig:ig}Influence of the number of generated news articles on the AUC metric over four base models.}
\end{figure}

\begin{table*}[t]
\renewcommand{\arraystretch}{1.3} 
\setlength\tabcolsep{3pt}
\caption{\label{tab:sm-full}Performance comparison on the MIND dataset, among the one-pass content summarizer (CS), one-pass user profiler (UP), one-pass personalized content generator (CG), chain-based personalized content generator (UP$\rightarrow$CG), and ALL that combines the content title generated by the one-pass content summarizer and synthetic content generated by the chain-based personalized content generator. ORI: training with the original data.}
\resizebox{\linewidth}{!}{
\begin{tabular}{cccccccccccccccccc}
\toprule[1pt]
\footnotesize{\textbf{Matching}} &  & \multicolumn{4}{c}{\textbf{NAML}} & \multicolumn{4}{c}{\textbf{LSTUR}} & \multicolumn{4}{c}{\textbf{NRMS}} & \multicolumn{4}{c}{\textbf{PLMNR}} \\
% \cmidrule(lr){2-5} \cmidrule(lr){6-9} \cmidrule(lr){10-13} \cmidrule(lr){14-17}
\cmidrule(lr){3-6} \cmidrule(lr){7-10} \cmidrule(lr){11-14} \cmidrule(lr){15-18}

& & \textbf{AUC} & \textbf{MRR} & \textbf{N@5} & \textbf{N@10} & \textbf{AUC} & \textbf{MRR} & \textbf{N@5} & \textbf{N@10} & \textbf{AUC} & \textbf{MRR} & \textbf{N@5} & \textbf{N@10} & \textbf{AUC} & \textbf{MRR} & \textbf{N@5} & \textbf{N@10} \\
\midrule
\textbf{ORI} & 
    & 61.75 & 30.60 & 31.35 & 37.85 
    & 61.27 & 29.64 & 30.28 & 36.76 
    & 61.71 & 30.20 & 30.98 & 37.42 
    & 62.53 & 30.74 & 31.31 & 38.03 \\
\textbf{CS} & 
    & 63.73 & 31.83 & 32.94 & 39.24
    & 62.16 & 30.52 & 31.27 & 37.85 
    & \textbf{63.85} & 31.57 & 32.35 & 38.80 
    & 64.80 & \textbf{33.08} & 34.25 & 40.35 \\
\textbf{UP} & 
    & 62.19 & 30.90 & 31.78 & 38.26 
    & 61.81 & 30.39 & 31.00 & 37.46 
    & 61.90 & 30.60 & 31.54 & 37.66 
    & 63.31 & 31.58 & 32.65 & 38.87 \\
\textbf{CG} & 
    & 62.93 & 30.83 & 32.10 & 38.34 
    & 63.88 & 31.76 & 32.92 & 39.16
    & 63.04 & 31.00 & 31.84 & 38.22 
    & 63.11 & 30.90 & 32.02 & 38.37 \\
\textbf{UP$\rightarrow$CG} & 
    & 63.61 & 31.58 & 32.63 & 39.07
    & 63.57 & 31.43 & 32.62 & 39.01
    & 62.95 & 32.00 & 32.80 & 39.00
    & 64.02 & 31.98 & 33.25 & 39.40
\\
\textbf{ALL} & 
    & \textbf{63.88} & \textbf{32.17} & \textbf{33.14} & \textbf{39.37}
    & \textbf{64.04} & \textbf{32.40} & \textbf{33.30} & \textbf{39.47}
    & 63.71 & \textbf{32.14} & \textbf{33.11} & \textbf{39.43} 
    & \textbf{65.13} & 32.98 & \textbf{34.30} & \textbf{40.49} \\
\midrule[1pt]
\footnotesize{\textbf{Ranking}} & & \multicolumn{4}{c}{\textbf{BST}} & \multicolumn{4}{c}{\textbf{DCN}} & \multicolumn{4}{c}{\textbf{PNN}} & \multicolumn{4}{c}{\textbf{DIN}} \\
% \cmidrule(lr){2-5} \cmidrule(lr){6-9} \cmidrule(lr){10-13} \cmidrule(lr){14-17}
\cmidrule(lr){3-6} \cmidrule(lr){7-10} \cmidrule(lr){11-14} \cmidrule(lr){15-18}
& & \textbf{AUC} & \textbf{MRR} & \textbf{N@5} & \textbf{N@10} & \textbf{AUC} & \textbf{MRR} & \textbf{N@5} & \textbf{N@10} & \textbf{AUC} & \textbf{MRR} & \textbf{N@5} & \textbf{N@10} & \textbf{AUC} & \textbf{MRR} & \textbf{N@5} & \textbf{N@10} \\
\midrule
\textbf{ORI} & 
    & 61.73 & 29.84 & 30.55 & 37.22
    & 62.63 & 29.73 & 30.52 & 37.12 
    & 61.75 & 29.45 & 29.99 & 36.67 
    & 60.95 & 28.13 & 28.77 & 35.42 \\
\textbf{CS} & 
    & 62.85 & 31.51 & 32.16 & 38.78
    & 64.19 & 31.96 & 32.67 & 39.16
    & 63.85 & 31.54 & 32.38 & 38.78
    & 61.26 & 29.72 & 30.38 & 36.76 \\
\textbf{UP} & 
    & 62.67 & 30.75 & 31.63 & 38.01
    & 63.47 & 29.92 & 30.66 & 37.47 
    & 62.34 & 29.67 & 30.46 & 37.07 
    & 62.65 & 30.74 & 31.50 & 38.05 \\
\textbf{CG} & 
    & 62.86 & 30.54 & 31.32 & 37.93 
    & 62.67 & 29.81 & 30.63 & 37.18 
    & 62.24 & 29.34 & 30.05 & 36.73 
    & 62.18 & 29.33 & 29.88 & 36.79  \\
\textbf{UP$\rightarrow$CG} & 
    & 63.28 & 31.49 & 32.45 & 38.84 
    & 63.05 & 29.79 & 30.61 & 37.23
    & 63.63 & 30.85 & 31.14 & 38.69
    & 63.53 & 30.76 & 31.21 & 38.13  \\
\textbf{ALL} & 
    & \textbf{63.94} & \textbf{32.05} & \textbf{33.09} & \textbf{39.41}
    & \textbf{65.77} & \textbf{32.86} & \textbf{34.10} & \textbf{40.48}
    & \textbf{65.49} & \textbf{32.78} & \textbf{33.81} & \textbf{40.19}
    & \textbf{63.80} & \textbf{31.68} & \textbf{32.57} & \textbf{39.08} \\
\bottomrule[1pt]
\end{tabular}
}
\end{table*}

\begin{table}[!h]
\renewcommand{\arraystretch}{1.2}
\setlength\tabcolsep{2pt}
\caption{\label{tab:cost}Comparison of the cost and cost conversion rate (CCR) of different generative schemes. Imp.: the average improvement in AUC compared with the original dataset. CCR: the ratio of improvement to cost. Note that the cost of $\text{UP}\rightarrow\text{NG}$ is calculated by $120 \times 0.21 + 60$, where $120$ is the cost of $\text{UP}$, $0.21$ is the new user ratio, and $60$ is the cost of chain-based $\text{NG}$.}
% \resizebox{\linewidth}{!}{
\begin{tabular}{ccccccc}
\toprule
 &  & \multicolumn{2}{c}{\textbf{Matching}} & \multicolumn{2}{c}{\textbf{Ranking}} \\
\cmidrule(lr){3-4} \cmidrule(lr){5-6}
 & \textbf{Cost} (USD) & \textbf{Imp.} & \textbf{CCR} (\%) & \textbf{Imp.} & \textbf{CCR} (\%) \\
\midrule
% \textbf{NS} & 60 & \textbf{6.78} & 11.3 & \textbf{6.87} & \textbf{11.5} \\
% \textbf{UP} & 120 & 2.61 & 2.18 & 4.18 & 3.48 \\
% \textbf{NG} & 40 & 4.74 & \textbf{11.9} & 2.25 & 5.63 \\
\textbf{NS} & 60 & \textbf{1.82} & 3.03 & \textbf{1.27} & \textbf{2.11} \\
\textbf{UP} & 120 & 0.49 & 0.41 & 1.02 & 0.85 \\
\textbf{NG} & 40 & 1.42 & \textbf{3.55} & 0.72 & 1.80 \\ \midrule
\textbf{NG} & 40 & 1.47 & \textbf{3.68} & 0.95 & \textbf{2.38} \\
\textbf{UP$\rightarrow$NG} & 85 & \textbf{2.43} & 2.86 & \textbf{1.57} & 1.85 \\
\bottomrule
\end{tabular}
% }
\end{table}

\subsection{Performance Comparison}

\autoref{tab:sm-full} presents the performance comparison for 1) one-pass content summarizer (CS), 2) one-pass user profiler (UP), and 3) one-pass personalized content generator (CG), 4) chain-based content generator (UP $\rightarrow$ CG) describe in the previous section, and 5) the combination of CS and UP $\rightarrow$ CG (ALL). The results in \autoref{tab:sm-full} show that: \textbf{Firstly,} the combination of the three generative schemes (i.e., ``ALL'') achieve the best performance for all recommendation models in most cases, significantly outperforming training with the original data (ORI). 
\textbf{Secondly,} chain-based personalized content generator performers better than one-pass variants, which indicates the effectiveness of such chain-of-thought prompt.

\newpage

\subsection{Content Summarizer}

The text feature in the Goodreads dataset is only the book title, while in the MIND dataset, there are news title, abstract, and category.
In the above experiments, we only utilize (enhanced) news title and category as inputs to the content encoder. Here, we assess the impact of combining more news features. 
From~\autoref{fig:fe-features}, the following can be summarized. \textbf{Firstly}, the inclusion of additional news features such as abstract and subcategory does lead to an improved model performance, although they are usually excluded from existing models out of efficiency concerns. 
\textbf{Secondly}, while MIND* has included all available news features, MIND-NS* still outperform MIND*, 
indicating the effectiveness of the news titles generated by GPT-3.5.

\subsection{Personalized Content Generator}

Here, we study how the number of generated content affects the recommendation performance. As depicted in~\autoref{fig:ig} conducted on the MIND dataset, we evaluate the effectiveness of utilizing 0, 1, and 2 generated news articles per new user for four base models. It can be seen that for each model, the performance improves as the number of generated content increases.

\subsection{Cost Conversion Rate}

Finally, we investigate the cost and cost conversion rate (CCR) of different generative schemes under our ONCE framework,
%the economic advantages of utilizing the three aspects of GENRE framework, 
as presented in~\autoref{tab:cost}. We compute the average improvement in AUC compared with the original dataset
%performance boost of the three enhanced datasets on 
for both matching and ranking models based on the results from~\autoref{tab:sm-full}, as well as the cost conversion rate (ratio of improvement in AUC to cost of employing the GPT-3.5 API). Based on the results, we can conclude the following. %\textbf{Firstly}, news summarizer and personalized news generator gain the best CCR in full dataset in the matching and ranking group, respectively. 
\textbf{Firstly}, with the full dataset, the personalized content generator (CG) has the best CCR for matching-based models, and the content summarizer (CS) has the best CCR for ranking-based models. \textbf{Secondly}, the user profiler (UP) has the worst CCR, since the extensive length of a user's browsing history results in a high token count per request, leading to increased cost for the user profiler. \textbf{Thirdly}, chain-based generation achieves a higher improvement compared to one-pass generation, but its CCR decreases due to the use of the expensive user profiler.

\end{document}